%% file: cmnlg.tex
\documentclass[11pt]{article}
\usepackage{jheppub}

\usepackage{amsmath}
\usepackage{amsfonts}
\usepackage{tikz}
\usepackage{graphicx}
\usepackage{dcolumn}
\usepackage{amssymb}
\usepackage{bbold}
\usepackage{comment}

\makeatletter
\renewcommand{\@seccntformat}[1]{\csname the#1\endcsname.\,\,}
\makeatother
\allowdisplaybreaks

\usepackage{xcolor}
\definecolor{vub}{RGB}{0,52,154}
\usepackage{xargs}
\usepackage[colorinlistoftodos,prependcaption,textsize=tiny]{todonotes}
\newcommandx{\unsure}[2][1=]{\todo[linecolor=vub,backgroundcolor=vub!25,bordercolor=vub,#1]{#2}}

\usepackage{float}
\usepackage[normalem]{ulem}

\usepackage{mathrsfs} 

\usepackage{multicol}
\usepackage{cancel}
\usepackage{mathtools}
\usepackage{hyperref}

\usepackage{makecell}
\usepackage{braket}

\newcommand\blfootnote[1]{%
\begingroup 
\renewcommand\thefootnote{}\footnote{#1}%
\addtocounter{footnote}{-1}%
\endgroup 
}

\def\({\left(} \def\){\right)}
\def\[{\left[} \def\]{\right]}

\newcommand{\eg}{\textit{e.g.}\,,\, }
\newcommand{\ie}{\textit{i.e.}\,,\, }

\def\tL{\tilde{L}}

\newcommand{\bea}{\begin{eqnarray}}
\newcommand{\eea}{\end{eqnarray}}
\def\p{\partial}
\usepackage{changepage}

\newcommand{\beq}{\begin{equation}}
\newcommand{\eeq}{\end{equation}}

\newcommand{\dd}{\mathrm{d}}

\def\le{\left(}
\def\ri{\right)}

\renewcommand{\eqref}[1]{(\ref{#1})}

\title{Conformal Mapping of Non-Lorentzian Geometries in SU(1\,,\,2) Conformal Field Theory}
\author{Stefano Baiguera$^{1, \, 2}$, Troels Harmark$^{3, \, 4}$, Yang Lei$^{5}$, and Ziqi Yan$^{4}$ \blfootnote{*The authors are ordered purely alphabetically and should all be viewed as the co-first authors. }}

\affiliation{$^1$INFN Sezione di Perugia, Via A. Pascoli, 06123 Perugia, Italy}
\affiliation{$^2$Dipartimento di Matematica e Fisica, Universit\`{a} Cattolica del Sacro Cuore, \\ Via della Garzetta 48, 25133 Brescia, Italy}
\affiliation{$^3$Niels Bohr International Academy, Niels Bohr Institute, University of Copenhagen, \\
Blegdamsvej 17, DK-2100 Copenhagen Ø, Denmark}
\affiliation{$^4$Nordita, KTH Royal Institute of Technology and Stockholm University, \\
Hannes Alfvéns väg 12, SE-106 91 Stockholm, Sweden}
\affiliation{$^5$Institute for Advanced Study \& School of Physical Science and Technology,\\
Soochow University, Suzhou 215006, P.R. China}

\emailAdd{stefano.baiguera@pg.infn.it}
\emailAdd{harmark@nbi.ku.dk}
\emailAdd{leiyang@suda.edu.cn}
\emailAdd{ziqi.yan@su.se}

\abstract{We realize an explicit conformal mapping between the state and operator pictures in a class of ($2+1$)-dimensional non-Lorentzian field theories with SU(1\,,\,2)$\times$U(1) 
conformal symmetry.  The state picture arises from null reducing four-dimensional relativistic conformal field theories on a three-sphere, yielding a non-Lorentzian geometry with the conformal Killing symmetry group SU(1\,,\,2). This is complementary to the operator picture recently studied by Lambert~\emph{et al.}~\cite{Lambert:2021nol}, where the geometry acquires an $\Omega$-deformation. We then use the geometric mapping between the two pictures to derive a correspondence between the generators. This provides a concrete realization of the state-operator correspondence in non-Lorentzian conformal field theories.} 

\preprint{NORDITA 2024-043}

\begin{document}

\maketitle

\setcounter{tocdepth}{1}

\input{Sections/Introduction}

\input{Sections/Review_relativistic}

\input{Sections/Nullred_geometries}

\input{Sections/Map}

\input{Sections/Discussion}

\acknowledgments

We would like to thank Christian Northe for valuable discussions.
The work of S.B. is supported by the INFN grant \textit{Gauge Theories and Strings (GAST)} via a research grant on \textit{Holographic dualities, quantum information and gravity}.
The work of S.B. was also supported by the Israel Science Foundation (grant No. 1417/21), by the German Research Foundation through a German-Israeli Project Cooperation (DIP) grant “Holography and the Swampland”, by Carole and Marcus Weinstein through the BGU Presidential Faculty Recruitment Fund, by the ISF Center of Excellence for theoretical high energy physics.
S.B. gratefully acknowledges the Niels Bohr Institute Academy (NBIA), the Nordic Institute for Theoretical Physics (Nordita) and the Oskar Klein Centre (OKC) for hospitality.
S.B. thanks the COST action (project \textit{Fundamental challenges in theoretical physics} CA22113) for the award of a short-term scientific mission at NBIA, where part of the research in this paper was carried out.
T.H. thanks Nordita for the support as corresponding fellow.
Y.L. is supported by a Project Funded by the Priority Academic Program Development of Jiangsu Higher Education Institutions (PAPD) and by National Natural Science Foundation of China No.12305081 and joint travel grant between National Natural Science Foundation of China (NSFC) and International Center for Theoretical Physics (ICTP) No.12481540178.
The work of Z.Y. is supported in
part by the European Union’s Horizon 2020 research and innovation programme under the
Marie Sklodowska-Curie Grant Agreement No. 31003710, VR Project Grant
2021-04013, and Olle Engkvists Stiftelse Project Grant 234-0342. Nordita is supported in part by NordForsk.

\newpage

\appendix

\input{Sections/App_su12algebra}

\input{Sections/App_details_map}

\bibliographystyle{JHEP}

\bibliography{cmnlg}

\end{document}

%% file: Sections/Introduction.tex
\section{Introduction}

There has been significant interest in non-Lorentzian systems that exhibit an anisotropic scaling symmetry between space and time. Such systems display interesting quantum critical phenomena and have a wide range of applications in both condensed matter and high energy physics \cite{coleman2005quantum,Sachdev_2011,Sachdev:2011cs,Hoyos:2013eza,Hoyos:2013qna,Chapman:2014hja,Horava:2009uw,Hartnoll:2009ns}.
In particular, this includes systems at a Lifshitz fixed point, where the dynamical critical exponent $z= 2$ is of special interest~\cite{Kachru:2008yh}, with time $t$ and spatial coordinates $\mathbf{x}$ scaling differently as $t \rightarrow b^2 \, t$ and $\mathbf{x} \rightarrow b \, \mathbf{x}$\,, respectively, for a constant scaling factor $b$\,. 
This unique scaling allows for extensions of the spacetime symmetry to include Galilean boosts and a special conformal transformation, which result in the Schr\"{o}dinger group that serves as a non-Lorentzian analogue of the relativistic conformal group (see \cite{Baiguera:2023fus,Duval:2024eod} for recent reviews).  

Quantum field theories (QFTs) invariant under the Schr\"{o}dinger group are referred to as Schr\"{o}dinger Conformal Field Theories (CFTs). Studies of such unconventional CFTs have triggered the program of non-Lorentzian bootstrap, for which some first steps are given in~\cite{Golkar:2014mwa,Goldberger:2014hca,Gubler:2015iva,Chen:2020vvn,Chen:2022jhx}. For instance, fermions at unitarity are governed by the Schr\"{o}dinger symmetry~\cite{Son:2005rv,Nishida:2006br,Nishida:2007pj,Son:2008ye,Nishida:2010tm,Bekaert:2011qd,Raviv-Moshe:2024yzt}, which have applications to scatterings in nuclear physics and ultra-cold atomic systems~\cite{Kaplan:1998tg,Bedaque:1998kg,Regal,Zwierlein}. 
Following an algebraic approach, it was shown that Schr\"{o}dinger-invariant theories admit a nonrelativistic version of the state-operator correspondence \cite{Nishida:2007pj}. 
In this context, primary operators are related to the eigenstates of a Hamiltonian with harmonic potential, and the corresponding eigenvalue is proportional to the scaling dimension of the local operator.
Moreover, Schr\"{o}dinger CFTs have also been playing an important role in bottom-up approaches to non-Lorentzian holography~\cite{Balasubramanian:2008dm,Son:2008ye}.

In this paper, we study a novel class of (2+1)-dimensional non-Lorentzian CFTs that is invariant under the SU$(1, 2) \times$U(1) conformal symmetry. This conformal group can be obtained by deforming the Schr\"{o}dinger group. The motivation for studying such an exotic symmetry group SU$(1, 2)$ stems from a top-down construction of non-Lorentzian holography. 
This endeavor is part of the recent program of deriving non-Lorentzian holography from decoupling limits of string theory and of the AdS/CFT correspondence (see \emph{e.g.}~\cite{Oling:2022fft,Baiguera:2023fus} for recent reviews). Such decoupling limits are incarnated as the closely related notions of near-BPS limits or null compactifications~\cite{Blair:2024aqz}. Intriguingly, in contrast to the bottom-up approach where the Schr\"{o}dinger symmetry prevails in the literature, the SU$(1, 2)$ group arises more naturally in the top-down approach. 

In the original proposal of the AdS/CFT correspondence, strongly-coupled $\mathcal{N} = 4$ SYM is mapped to weakly-coupled gravity on AdS${}_5 \times S^5$\,. One prominent way to further deepen our understanding of this correspondence is by zooming in on self-consistent sub-sectors of $\mathcal{N} = 4$ SYM, where there is a better chance to obtain exact field-theoretical results that can be mapped to the gravity side. This philosophy has been pursued since the early days of the AdS/CFT correspondence, either by working in the integrable regime at infinite $N$~\cite{Minahan:2002ve,Beisert:2003tq,Beisert:2003ys,Beisert:2004ry,Bellucci:2005vq,Bellucci:2006bv,Beisert:2007sk,Zwiebel:2007cpa,Beisert:2008qy,Beisert:2010jr}, or by considering the Berenstein-Maldacena-Nastase (BMN) and Penrose limits~\cite{Berenstein:2002jq,Kruczenski:2003gt}. In these settings, $1/N$ contributions only enter perturbatively, making it difficult to access non-perturbative regimes on the gravity side.
In contrast, it is shown in~\cite{Harmark:2014mpa} that the Spin Matrix Theory (SMT) limit on the field theory side, where $N$ is kept finite, allows us to go beyond this perturbative treatment. 

One common feature of the BMN, Penrose, and SMT limits of the AdS/CFT correspondence is that a null isometry is introduced on the bulk side. Consequently, the bulk null isometry induces a near-BPS decoupling limit on the field theory side, such that the particle number becomes conserved. Such quantum mechanical systems may also be viewed as non-Lorentzian CFTs, where the spacetime is anisotropic and typically develops Galilei-like boost symmetry. Nevertheless, the formulation of these non-Lorentzian CFTs is much less developed than the quantum mechanical approach, which clearly is a hurdle against a comprehensive understanding of how the analogous holographic principle works for these corners of $\mathcal{N} = 4$ SYM.   

Recently, intriguing progress along these lines was made by Lambert, Mouland, and Orchard~in~\cite{Lambert:2021nol}, where ($2n$-1)-dimensional non-Lorentzian CFTs with an exotic SU$(1,n)$ spacetime symmetry were introduced.\footnote{Geometries with SU$(1,n)$ symmetry naturally arise from a top-down approach to non-Lorentzian holography in~\emph{e.g.}~\cite{Lambert:2019jwi, Lambert:2019fne,Lambert:2020jjm,Lambert:2020zdc,Lambert:2024buc}.} In this proposal, the $z=2$ Lifshitz scaling and Galilean boosts are realized in a non-trivial way. The operator picture of SU$(1\,, \, n)$ CFTs was obtained from a conformal compactification of a $2n$-dimensional Lorentzian CFT in Minkowski spacetime, leading to an $(2n-1)$-dimensional non-Lorentzian geometry. However, the question of what the non-Lorentzian spacetime geometry should be in the state picture still remains open. Moreover, it is important to look for an explicit conformal mapping between the non-Lorentzian geometries in the state and operator pictures. Answers to these questions would provide important geometric intuitions about the state-operator correspondence in non-Lorentzian CFTs, which so far has been mostly studied algebraically~\cite{Nishida:2007pj,Lambert:2021nol,Karananas:2021bqw}. 

\vskip 2mm

In this work, we will tackle these questions for the specific case of $n = 2$\,, corresponding to a (2+1)-dimensional CFTs with an SU($1,2$) symmetry. This example is of particular interest due to its connection to $\mathcal{N} = 4$ SYM in four dimensions, which is invariant under the conformal group $\mathrm{SO}(2,4) \approx \mathrm{SU}(2,2)$\,. We will begin by constructing the state picture via a null reduction of the Lorentzian cylinder $\mathbb{R} \times S^3$\,. Such a null reduction leads to a non-Lorentzian geometry in one lower dimension, which does \emph{not} admit any (2+1)-dimensional metric description. Instead, this non-Lorentzian geometry is encoded by vielbein fields that form the so-called \emph{torsional Newton-Cartan} (TNC) geometry~\cite{Christensen:2013lma, Christensen:2013rfa}, which generalizes the Newton-Cartan framework that covariantizes the spacetime structure underlying Newtonian gravity. In a TNC geometry, the strict notion of absolute time inherent in Newton-Cartan geometry is relaxed. We will then identify a timelike coordinate within this TNC geometry corresponding to the Hamiltonian evolution in the state picture of the non-Lorentzian CFT, which is used to build the Hilbert space. 
This non-Lorentzian CFT admits an SU$(1, 2) \times$U(1) symmetry, where SU$(1, 2)$ is the conformal Killing group of the (2+1)-dimensional TNC geometry, and U(1) corresponds to the isometry of the extra null direction when the geometry is uplifted back to four dimensions. 

Next, we will derive the operator-picture TNC geometry by sending the characteristic length scale in the state-picture TNC geometry to infinity. The resulting background reproduces the one in~\cite{Lambert:2021nol}, which arises from a null reduction of Minkowski spacetime $\mathbb{R}^{1\,, \, 3}$ with an $\Omega$-deformation. Finally, we will derive the conformal mapping between the state- and operator-picture TNC geometries, and compute how the Hamiltonian generating the evolution between equal-time slices in the state picture is mapped into a linear combination of the generators in the operator picture.  

This study may be viewed as a precursor for understanding the field-theoretical aspects of the SMT limits of $\mathcal{N} = 4$ SYM. Any SMT limit preserves a subgroup of the original $\mathrm{PSU}(2,2|4)$ symmetry of $\mathcal{N}=4$ SYM \cite{Harmark:2007px}.
The most significant SMT is the $\mathrm{PSU}(1,2|3) \times \mathrm{U}(1)$ sector, which is the largest subgroup, allowing one to recover all the other cases by constraining its field content. The additional $\mathrm{U}(1)$ factor is emergent in the SMT limit, and corresponds to the particle number conservation in a quantum mechanical system. The SMT Hamiltonians have been derived for all the sectors in~\cite{Harmark:2019zkn,Baiguera:2020jgy,Baiguera:2020mgk,Baiguera:2021hky,Baiguera:2022pll}.
The results are formulated in terms of a quantum mechanical Hamiltonian obtained by expanding in modes over the three-sphere.
Despite a QFT formulation for the SMTs with SU($1,1$) invariance exists in~\cite{Baiguera:2020jgy}, an off-shell local action for the general case is still missing.
The bosonic sector of the desired field-theoretical formulation of the SMT with PSU($1,2 | 3$) invariance must be a non-Lorentzian CFT with the SU($1,2$) symmetry, with natural connections to the model that we focus on in the current paper.

The paper is organized as follows.
We begin by reviewing in Section~\ref{sec:rel_state_ope} the state-operator correspondence in Lorentzian CFTs.
While this framework is typically more intuitive in Euclidean signature, we instead focus on its realization in real time (Lorentzian signature) due to its closer parallels with the non-Lorentzian case that we discuss in the rest of the paper.
Section~\ref{sec:cft_map_su12} forms the core of the paper.
Here, we construct the non-Lorentzian geometries in both the state and operator pictures via null reductions, each exhibiting SU($1,2$) conformal isometry.
We then establish a conformal mapping between these backgrounds, which allows us to enhance the formulation of the state-operator correspondence in non-Lorentzian CFTs. The detailed derivation of the conformal map is given in Section~\ref{sec:state_ope_map}. 
The summary of results and a discussion of the relevant applications in a broader context, including holography, are presented in Section~\ref{sec:discussions}.
Additional technical details are collected in Appendix~\ref{app:su12_algebra} and Appendix~\ref{app:ssec:coord_ope}, regarding $\mathfrak{su}(1,2)$ algebra and coordinate transformations for the operator picture, respectively.

%% file: Sections/Review_relativistic.tex
\section{Review: Relativistic State-Operator Correspondence}
\label{sec:rel_state_ope}

Due to the simplicity of the conformal map between the Euclidean plane and the cylinder, the state-operator correspondence is usually formulated in imaginary time with Euclidean signature. 
In contrast, non-trivial features arise when the correspondence is studied in Lorentzian signature (real time), which are seldomly discussed in the literature.
As we aim to construct a conformal map in non-Lorentzian CFTs from performing certain null reductions, it is more natural for us to use the real-time formulation. Before studying non-Lorentzian CFTs, we first review some salient features of the relativistic state-operator correspondence in both Euclidean and Lorentzian signatures.

\subsection{Euclidean Time}
\label{ssec:eucl_time}

The state-operator correspondence represents an elegant isomorphism in CFT between the space of states on a ($d-1$)-dimensional sphere and the set of local operators in $d$-dimensional Minkowski spacetime~\cite{DiFrancesco:1997nk,Rychkov:2016iqz,Simmons-Duffin:2016gjk}. This identification requires first compactifying all the spatial directions on an $S^{d-1}$ in the state picture. In a relativistic theory where time and space have the same scaling dimension, it is convenient to perform a Wick rotation, such that the Minkowski spacetime becomes a Euclidean plane in the operator picture.
Then, it is simple to realize a bijectiive map between the cylinder $\mathbb{R} \times S^{d-1}$ in the state picture and the Euclidean plane $\mathbb{R}^d$ in the operator picture, by performing a conformal mapping. Denote the number of spatial dimensions by $d-1$\,, the Euclidean time in the state picture as $\tau$\,, and the radial coordinate in the operator picture as $\rho$\,.  
Up to a conformal factor, the desired bijective map between the line elements is given by
\definecolor{dblue}{RGB}{0, 0, 180}
\begin{align} \label{eq:relsoc}
\begin{split}
& \hspace{3mm} \text{\emph{state picture}} 
    \hspace{3.8cm}%
\text{\emph{operator picture}} \\[2pt]
\dd s^2 & = e^{2\tau} \le \dd \tau^2 + \dd \Omega_{d-1}^2 \ri
\,\,\quad\overset{\tau = \ln \rho \,}{\xrightarrow{\hspace{1cm}}}\quad\,\,         
\dd s^2 =  \dd \rho^2 + \rho^2 \, \dd \Omega_{d-1}^2  \\[10pt]
&\begin{minipage}{10.7cm}
\begin{tikzpicture}[scale=0.48]
\draw[very thick] (-2-2.5, 0) to  (-2-2.5, 6);
\draw[very thick] (2-2.5, 0) to  (2-2.5, 6);
\draw[very thick] (0-2.5, 0.1) ellipse (2cm and 0.6cm);
\draw[very thick] (0-2.5,6) ellipse (2cm and 0.5cm);
\draw[dashed, very thick] (0-2.5,1.5) ellipse (2cm and 0.5cm);
\draw[dashed, very thick] (0-2.5,3) ellipse (2cm and 0.5cm);
\draw[dashed, very thick] (0-2.5,4.5) ellipse (2cm and 0.5cm);
\fill[white] (-1.95-2.5, 0) rectangle (1.96-2.5, 0.8);
\fill[white] (-1.95-2.5, 1.5) rectangle (1.96-2.5, 2.1);
\fill[white] (-1.95-2.5, 1.5 + 1.5) rectangle (1.96-2.5, 2.1 + 1.5);
\fill[white] (-1.95-2.5, 1.5 + 3) rectangle (1.96-2.5, 2.1 + 3);
\draw[very thick, ->, dblue] (-2.7-2.5, 1.5) to  (-2.7-2.5, 4.5);
\node at (-3.2-2.5,3) {$\tau$};
\draw[very thick, ->] (3-1.25, 3) -- (6-1.25, 3);
\node at (4.5-1.25, 3.6) {\scalebox{0.8}{$\tau = \ln \rho$}};
\draw[->,ultra thick] (7,3)--(15,3) node[right]{$x$};
\draw[->,ultra thick] (11,-1)--(11,7) node[above]{$y$};
\draw[very thick, dashed] (11,3) circle (0.8cm);
\draw[very thick, dashed] (11,3) circle (1.6cm);
\draw[very thick, dashed] (11,3) circle (2.4cm);
\draw[very thick, ->, dblue] (11, 3) to  (13.8, 4.8); 
\node at (14, 5.2) {$\rho$};
\end{tikzpicture}
\end{minipage} \\
\end{split}
\end{align}
where an additional Weyl transformation can be performed to get rid of the overall $e^{2\tau}$ prefactor in front of the metric in the state picture. 
This conformal transformation then induces a map between generators in both pictures. In particular, the Hamiltonian $H^\text{E}$ on the cylinder is mapped to the dilatation operator $D^\text{E}$ on the Euclidean plane, 
\begin{equation}
    H^\text{E} = \p_\tau
    \,\,\overset{\tau = \ln \rho}{\xrightarrow{\hspace{1cm}}}\,\,         
    D^\text{E} = \rho \, \p_\rho\,.
    \label{eq:Euclidean_state_ope}
\end{equation}
Here, the superscript ``E'' stands for ``Euclidean.'' 
A state prepared at $\tau = - \infty$ on the cylinder is then mapped to a local operator with definite scaling dimension at the origin of the Euclidean plane at $\rho = 0$\,, while the Hamiltonian time evolution of this initial asymptotic state is mapped to an evolution along the radial direction governed by the dilatation operator.\footnote{This step is possible because the conformal transformation relates the Euclidean cylinder to a compact manifold (here it is the Euclidean plane), so that the non-local boundary condition at $\tau=-\infty$ on the cylinder maps to a local condition at a single point. This observation was stressed in \cite{Belin:2018jtf}, where the possible extension of the state-operator correspondence to toric geometries was investigated.}
Equal-time slices on the cylinder correspond to a radial foliation on the Euclidean plane.
This profound, yet relatively simple, realization of the state-operator correspondence plays a pivotal role in all the studies of CFTs. 

Needless to say, the conformal mapping~\eqref{eq:relsoc} only acquires a simple geometric interpretation when the space and time directions share the same scaling dimension. This is crucial for the equal-time hypersurfaces in the state picture to be mapped to concentric spheres in the operator picture. In contrast, in nonrelativistic systems where the Lorentz symmetry is absent, time and space can develop anisotropic scalings. In the case where certain non-Lorentzian boosts, such as Galilean or Carrollian, are present, the spacetime symmetry group can also be enhanced to include a non-Lorentzian analogue of the special conformal transformations. While the state picture 
can still be applied to nonrelativistic CFTs, one should not expect that the equal-time hypersurfaces are mapped to the concentric spheres in the operator picture anymore. In fact, the mapping of the equal-time slices to the operator picture must be represented by a polynomial function adapted to the higher-order dispersion relation of the degrees of freedom in the nonrelativistic CFT.  

Intriguingly, the geometric aspect of the state-operator correspondence in nonrelativistic CFTs is conceptually very similar to the correspondence in relativistic CFT in real time. Without performing the Wick rotation in the state picture, an equal-time hypersuface is mapped to a hyperbolic space in the operator picture. This real-time incarnation of the state-operator correspondence only makes calculations unnecessarily intricate in relativistic CFTs, but it provides essential intuitions for us to understand nonrelativistic CFTs. Below, we review some essential aspects of the relativistic state-operator correspondence in real time with a Lorentzian signature. We mainly follow~\cite{Minwalla:1997ka,Luscher:1974ez,Karananas:2021bqw} and Appendix~A of \cite{Chagnet:2021uvi}.

\subsection{Lorentzian Time} 
\label{sec:lt}

In real time, the geometric mapping in the state-operator correspondence is between the Einstein universe $\mathbb{R} \times S^{d-1}$ (which we will refer to as the Lorentzian cylinder) in the state picture and Minkowski spacetime $\mathbb{R}^{d-1,\,1}$ in the operator picture. The coordinate transformation relating these two manifolds is the hyperbolic map, which is only identified with the logarithmic map in Eq.~\eqref{eq:relsoc} via an analytic continuation of time. We start with the operator picture, and express Minkowski spacetime in spherical coordinates $x^\mu = (t\,, \, r\,, \, \cdots)$\,,
\begin{equation}
\label{mink_sphe}
    \dd s^2 = - \dd t^2 + \dd r^2 + r^2 \, \dd\Omega_{d-2}^{\, 2}\,.
\end{equation}
We perform the following coordinate transformation:
\beq
    t = \frac{1}{2} \left[  \tan \le \frac{\tau+\chi}{2} \ri + \tan \le \frac{\tau-\chi}{2} \ri \right],  
        \qquad%
    r = \frac{1}{2} \left[  \tan \le \frac{\tau+\chi}{2} \ri - \tan \le \frac{\tau-\chi}{2} \ri \right],
\label{eq:ctrelLor}
\eeq
where $\tau$ and $\chi$ are respectively the time and spatial radial direction in the state picture. The corresponding line element is then
\begin{equation}
    \dd s^2 = \bigl(\cos \tau + \cos \chi \bigr)^{-2} \Big(  - \dd\tau^2 + \dd\chi^2   + \sin^2 \chi \, \dd\Omega_{d-2}^{\, 2} \Big) \, .
\label{eq:conmappedMink}
\end{equation}
After performing a Weyl rescaling to get rid of the conformal factor, we precisely obtain the metric of the Lorentzian cylinder.
Note that $\tau \in [-\pi,\pi]$ and $\chi \in [0,\pi]$\,, \ie we only have access to a patch on the Lorentzian cylinder, in addition to the usual range of angular coordinates on the transverse sphere $S^{d-2}$.

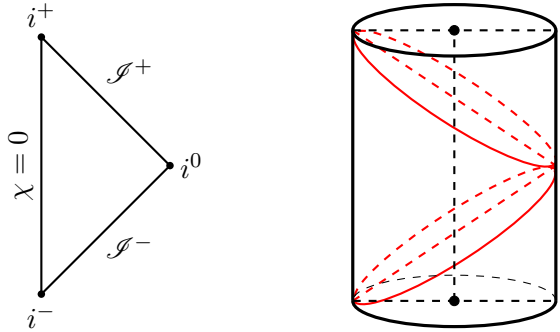
\begin{figure}[ht]
\centering
\tikzstyle{background rectangle}=[fill=blue!20]
\begin{tikzpicture}[scale=0.9]
\draw[red, thick, rotate=-33, dashed] (7.94,-5.2+10.4) arc(0:180:1.76cm and 0.33cm);
\draw[red, thick, rotate=-33] (4.4,-5.2+10.4) arc(180:360:1.77cm and 0.38cm);
\draw[thick,red,dashed] (6.5,2) to  (9.5,0);

\draw[red, thick, rotate=33, dashed] (7.93,-5.22) arc(0:180:1.76cm and 0.33cm);
\draw[red, thick, rotate=33] (4.39-0.02,-5.22) arc(180:360:1.77cm and 0.38cm);
\draw[thick,red,dashed] (6.5,-2) to  (9.5,0);

 \node[rotate=90] at (1.6,0) {$\chi=0$};
 \draw[thick] (1.9,1.9)--(1.9,-1.9);
 \draw[thick] (3.8,0) to (1.9,1.9);
 \draw[thick] (3.8,0) to (1.9,-1.9);
 \node[] at (3.2,-1.2) {$\mathscr{I}^-$};
 \node[] at (3.2,1.4) {$\mathscr{I}^+$};
 \node[below] at (1.9,-1.9) {$i^-$};
 \node[above] at (1.9,1.9) {$i^+$};
 \node[right] at (3.8,0) {$i^0$};
 \node at (1.9,-1.9) [circle,fill,inner sep=1pt]{};
 \node at (1.9,1.9) [circle,fill,inner sep=1pt]{};
 \node at (3.8,0) [circle,fill,inner sep=1pt]{};

\draw[dashed] (9.5,-2) arc(0:180:1.5cm and 0.38cm);
\draw[very thick] (6.5,-2) arc(180:360:1.5cm and 0.38cm);

\draw[very thick] (8,2) ellipse (1.5cm and 0.38cm);

\filldraw (8,2) circle (2pt); 
\filldraw (8,-2) circle (2pt); 

\draw[very thick] (6.5,-2) to  (6.5,2);

\draw[thick,dashed] (8,-2) to  (8,2);

\draw[very thick] (9.5,-2) to  (9.5,2);
\draw[thick,dashed] (6.5,-2) to  (9.5,-2);
\draw[thick,dashed] (6.5,2) to  (9.5,2);
\end{tikzpicture}
\caption{Left: Penrose diagram of Minkowski space after the compactification in Eq.~\eqref{eq:conmappedMink}. 
The points $i^{\pm}$ correspond to future (past) timelike infinity, while $i^0$ is spacelike infinity. 
$\mathscr{I}^{\pm}$ correspond to future (past) null infinity.
Right: the red lines delimit the patch of the Lorentzian cylinder obtained from Minkowski space via the conformal transformation corresponding to Eq.~\eqref{eq:ctrelLor}. 
 }
\label{fig:Penrose_Minkowski}
\end{figure}

The coordinate transformation leading to Eq.~\eqref{eq:ctrelLor} 
is obtained by introducing null coordinates and then compactifying them, a standard procedure that is used in textbooks to find the Penrose diagram of Minkowski space \cite{Carroll:2004st}, as depicted in fig.~\ref{fig:Penrose_Minkowski}.
The null boundaries of the conformal diagram correspond to $\mathscr{I}^{\pm}$, which delimit the portion of Lorentzian cylinder that is identified through the map \eqref{eq:ctrelLor} of Minkowski space.
On these surfaces, three special points are identified by timelike infinities $i^{\pm}$ (located at $\tau = \pm \pi$ and $\chi=0$) and spacelike infinity $i^0$ (located at $\tau=0$ and $\chi=\pi$).
These loci of spacetime are really \textit{points}, since they correspond to the north and south poles of the unit sphere $S^{d-1}$.

\begin{figure}[t!]
    \centering
    \hspace{1cm}\includegraphics[width=0.5\linewidth]{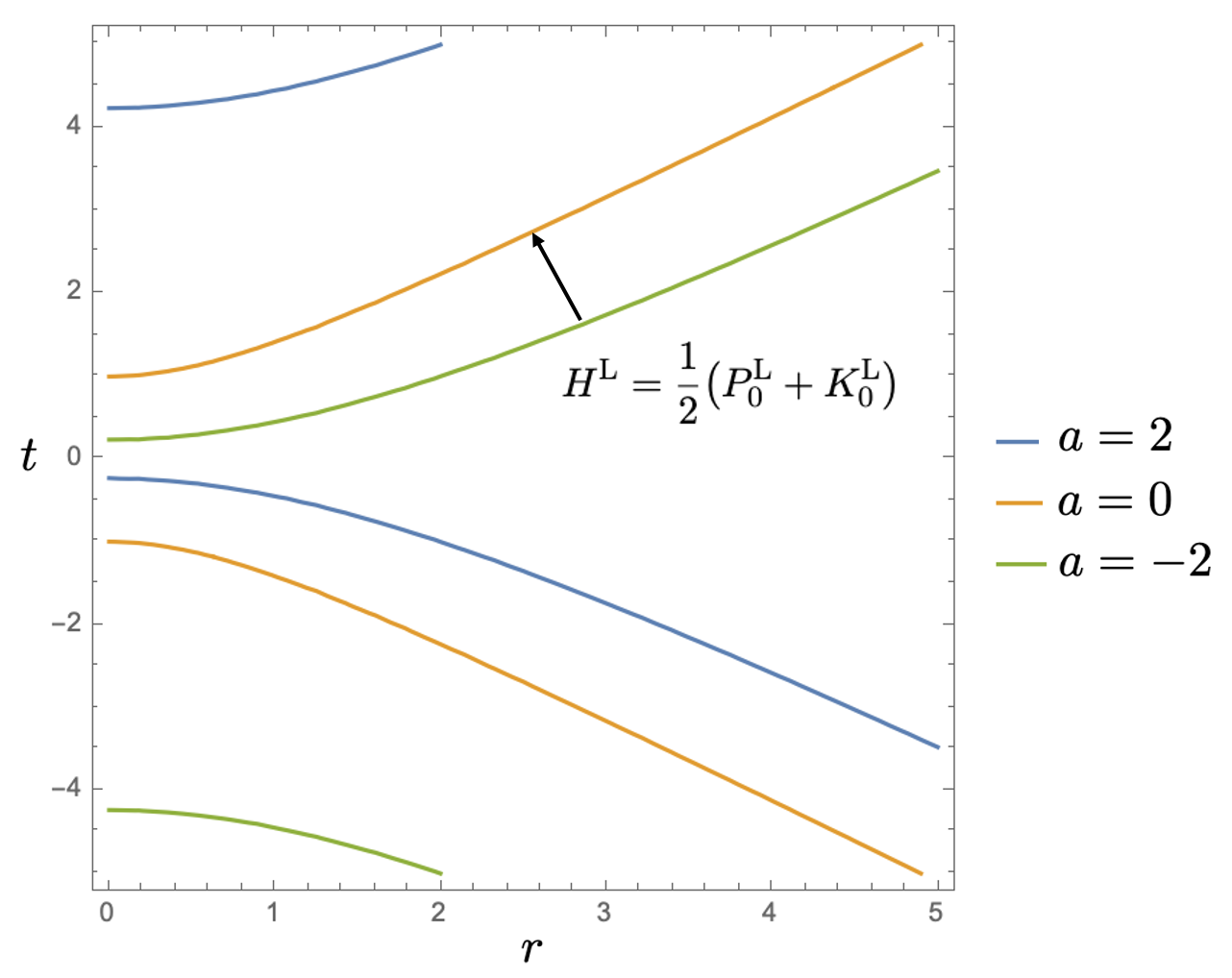}
    \caption{Plot of the hyperbolae \eqref{eq:hyperbolae_rel_case} in the $(t,r)$ plane.}
    \label{fig:hyp_rel}
\end{figure}

We now discuss how the generators transform under the coordinate change~\eqref{eq:ctrelLor} from Minkowski spacetime to the Lorentzian cylinder.
The Hilbert space on the Lorentzian cylinder in the state picture is constructed out of the eigenstates of the Hamiltonian
\bea
    H^\text{L} = \p_{\tau}\,,
\eea
which evolves states between different equal-time slices through its action. Here, ``L'' stands for ``Lorentzian.'' Using the inverse of the coordinate transformation~\eqref{eq:ctrelLor}, we find
\beq
    H^{\text{L}} = 
   \frac{1}{2} \bigl( 1 + t^2 + r^2 \bigr) \, \p_t +  \, t \, r \, \p_r\,.
\label{eq:relation_Htau_Lor}
\eeq
In terms of the momentum and special conformal generators associated with the conformal algebra in the operator picture,\,\footnote{See Eq.~\eqref{eq:diff_representation_conf_gen} for our conventions on the differential representation of the generators of the $\mathfrak{so}(d,2)$ algebra.}
\beq
     P^\text{L}_\mu = \frac{\p}{\p x^\mu}\,,
        \qquad%
     K^\text{L}_\mu =  x^{}_\nu \, x^\nu \, \frac{\p}{\p x^\mu} - 2 \, x^{}_\mu \, x^\nu \, \frac{\p}{\p x^\nu}  \,,
\eeq
we re-express the state-picture Hamiltonian $H^{\text{L}}$ in Eq.~\eqref{eq:relation_Htau_Lor} in terms of the operator-picture generators $P^{\text{L}}_0$ and $K^\text{L}_0$ as
\beq \label{eq:hpk}
    H^{\text{L}} = \frac{1}{2} \bigl( P_0^{\text{L}} + K_0^{\text{L}} \bigr)\,.
\eeq
Note that the corresponding generator in the operator picture is \emph{not} the dilatation operator $D = x^\mu \, \partial_\mu$ in Minkowski spacetime, which is not surprising as the equal-time slices in the state picture are mapped to hyperbolic surfaces that are not concentric as in the Euclidean case. To be specific, geometrically, constant $\tau$ slices on the Lorentzian cylinder are now mapped in Minkowski spacetime to the hyperbolae
\beq
     \frac{(t-a)^2}{1+a^2} - \frac{r^2}{1+a^2} = 1  \, ,
\label{eq:hyperbolae_rel_case}
\eeq
where $a=-\tan \tau$ is an integration constant at fixed $\tau$.
The operator $P^{\text{L}}_0 + K^{\text{L}}_0$ generates the evolution orthogonal to the hypersurfaces defined by Eq.~\eqref{eq:hyperbolae_rel_case}, which defines the relevant foliation used to build the Hilbert space in Minkowski space. 
The curves defining the foliation are depicted, for various choices of $a$, in Fig.~\ref{fig:hyp_rel}.
In the operator picture, timelike infinities $i^{\pm}$ are mapped to the spacelike slices at $t= \pm \infty$, corresponding to a degenerate limit of the hyperbolae where $a \rightarrow \pm \infty$, with $t/a \rightarrow \infty$.

Nevertheless, one would still expect that Eq.~\eqref{eq:hpk} can be identified with the dilatation operator in the operator picture after performing an analytic continuation to the imaginary time. 
By performing a similarity transformation (modulo an analytic continuation) on the generators in Lorentzian signature, it is possible to find a set of anti-Hermitian generators that provide a finite-dimensional and unitary representation of the $\mathfrak{so}(2\,,\,d)$ conformal algebra.\footnote{According to Wigner's theorem, any continuous symmetry connected to the identity and acting on a physical system must be represented by anti-Hermitian operators.}
They are used to build the Hilbert space by acting on local operators.
The key identity in the automorphism defines the Euclidean dilatation operator as
\beq
    D^\text{E} \equiv -\frac{i}{2} \bigl( P_0^\text{L} + K_0^\text{L} \bigr)\,,
        \qquad%
    H^\text{E} \equiv  -i H^\text{L}\,,
    \label{eq:rel_automorphism}
\eeq
and the state-operator correspondence maps from $H^\text{E}$ to $D^\text{E}$, as shown in Eq.~\eqref{eq:relation_Htau_Lor}.
This is the Lorentzian counterpart of the statement \eqref{eq:Euclidean_state_ope}, that was derived from an analysis in Euclidean signature only.

%% file: Sections/Nullred_geometries.tex
\section{Conformal Mapping of SU(\texorpdfstring{$1\,,\,2$}{1,2}) CFT}
\label{sec:cft_map_su12}

In this section, we consider an analogue of the geometric mapping reviewed in Section~\ref{sec:lt} for CFTs in Lorentzian time, but now in three-dimensional non-Lorentzian CFTs with Galilean symmetries, obtainable  by certain null reductions of relativistic CFTs.\,\footnote{Null reducing relativistic QFTs is a typical technique in the literature used to generate non-Lorentzian QFTs, where a relativistic theory compactified over a null circle and with a fixed null momentum corresponds to a Galilean-invariant theory. Galilean-invariant field theories generated with this mechanism have received a lot of attention in recent years, see,~\emph{e.g.},~\cite{Son:2008ye,Jensen:2014hqa,Jensen:2014aia,Auzzi:2015fgg,Auzzi:2016lrq,Auzzi:2016lxb,Arav:2016xjc,Festuccia:2016caf,Auzzi:2017jry,Auzzi:2017wwc,Auzzi:2019kdd,Lambert:2019jwi,Lambert:2019fne,Baiguera:2020jgy,Chapman:2020vtn,Lambert:2020jjm,Lambert:2020zdc,Lambert:2021nol,Bagchi:2022twx,Baiguera:2022cbp,Smith:2023jjb}.} An explicit realization of the coordinate transformation~\eqref{eq:ctrelLor} in non-Lorentzian CFT will allow us to propose a concrete state-operator correspondence in this context, which would serve as a foundation for building non-Lorentzian holography in the future. 
Whereas in Section~\ref{sec:lt} we considered general $d$, here we specialize to the case $d=4$ for which the Lorentzian geometric mapping is between the Lorentzian cylinder $\mathbb{R} \times S^3$ and four-dimensional Minkowski space.

The conformal group on four-dimensional Minkowski space is  $\mathrm{SU}(2\,,\,2)\approx \mathrm{SO}(2\,,\,4)$.
In the state picture, the metric on the Lorentzian cylinder enjoys a set of conformal Killing vectors (CKVs) that generate the same SU($2\,, \, 2$) group. 
It turns out that it is easier to 
make a connection to nonrelativistic CFTs by performing a null-reduction in the  
state picture, as compared to the operator picture, since the energy 
has a simpler definition in terms of the Killing vectors.\,\footnote{We will find that this is indeed the case later in Eq.~\eqref{eq:H0_ope}.} 

That it is beneficial to start in  the state picture is 
further
motivated by the AdS${}_5$/CFT${}_4$ correspondence, where $\mathcal{N} = 4$ SYM is dual to string theory on the bulk AdS geometry. In global coordinates, the bulk geometry approaches the background $\mathbb{R} \times S^3$ asymptotically. 
We have discussed in the Introduction that many important limits (BMN, Penrose, SMT, \emph{etc.}) of the AdS/CFT correspondence require a null isometry in the bulk geometry, which is usually only formulated using the global coordinates. At a fixed null momentum, this bulk null isometry naturally maps to a null reduction of the dual CFT in the state picture. 

Our goal is to build a state-operator correspondence in non-Lorentzian CFTs by starting with a relativistic CFT in the state picture, then perform a null reduction, and subsequently map out a relation to the corresponding operator picture. 
This construction will also have direct impact on holography, as we will later discuss in Section~\ref{sec:discussions}. We start with focusing on the geometric aspects of the state picture in Section~\ref{sec:nullred_RS3}, where we perform a null reduction to derive the three-dimensional TNC geometry for nonrelativistic CFTs in analogy with the Lorentzian cylinder~\eqref{eq:conmappedMink}. We find that such a null reduction leads to the conformal group of SU(1\,,\,2), which acquires a U(1) extension due to the extra null circle. We analyze the $\mathfrak{su}(1\,,\,2) \oplus \mathfrak{u} (1)$ algebra in Section~\ref{ssec:su12_algebra}.
We then determine in Section~\ref{ssec:infinite_radius_state} a natural background for the operator picture by performing an appropriate limit of the radius of the three-sphere in the state picture. The background obtained in this way is described by a TNC geometry with the same SU(1\,,\,2) conformal isometry as the state picture, as we show by relating the two metrics via a conformal transformation in Section~\ref{ssec:conf_map_NCgeo}.
This induces a map between the generators, that we study in Section~\ref{ssec:map_generators}.

\subsection{Null Reduction of \texorpdfstring{$\mathbb{R} \times S^3$}{RtimesS3} in State Picture}
\label{sec:nullred_RS3}

We now construct a null reduction on the $\mathbb{R} \times S^3$ Lorentzian cylinder in the state picture of a relativistic CFT. 
Stripping off the conformal factor, the $\mathbb{R} \times S^3$ line element in~\eqref{eq:conmappedMink} reads 
\beq \label{eq:dsdsr}
    \dd s^2 = - \dd \tau^2 + R^2 \, \dd\Omega^2_3\,,
        \qquad%
   \dd \Omega^2_3 = \dd \psi^2 + \cos^2 \psi \, \dd\phi^2_1 + \sin^2 \psi \, \dd\phi^2_2 \,,
\eeq
where $R$ is the radius of the three-sphere. Moreover, $\psi \in [0\,,\, \pi/2]$\,, $\phi_1\,, \, \phi_2 \in [0\,, \, 2\pi]$\,. 
It is convenient to perform the change of variables,
\beq 
    \theta = 2 \, \psi\,,
        \qquad%
    \eta = \phi^{}_1 + \phi^{}_2\,,
        \qquad%
    \varphi = \phi^{}_1 - \phi^{}_2\,,
\eeq
such that the line element~\eqref{eq:dsdsr} becomes
\beq \label{eq:lefdlc}
    \dd s^2 = - \dd \tau^2 + \tfrac{1}{4} \, R^2 \, \Bigl[ \dd \theta^2 + \sin^2 \theta \, \dd \varphi^2 + \bigl( \dd\eta + \cos \theta \, \dd \varphi \bigr)^2 \Bigr]\,,
\eeq
with $\theta \in [0\,, \, \pi]$\,, $\varphi \in [0\,, \, 2\pi]$\,, and $\eta \in [0\,,\, 4\pi]$\,. 
The conserved charges include the energy $E$ and angular momenta $S_1$ and $S_2$\,, with
\beq \label{eq:state_charges}
    E =  R \, \p^{}_{\tau}\,,
        \qquad%
    S^{}_1 = -  \, \p^{}_{\phi_1}\,,
        \qquad%
    S^{}_2 = -  \, \p^{}_{\phi_2}\,,
        \qquad%
    S \equiv S^{}_1 + S^{}_2 = - 2 \,  \, \p^{}_\eta\,,
\eeq
where the energy is measured in units of $R^{-1}$ and $S$ is the total angular momentum. 

In order to identify a non-Lorentzian CFT, we now define a null vector $\p_u$ by mixing the conserved charges $E$ and $S$ in \eqref{eq:state_charges}, such that 
\begin{equation} \label{eq:nullv}
    - \p_u \equiv E - S\,.
\end{equation}
This choice is partly motivated by the fact that the stability of the system requires that we restrict to the near-BPS excitations. Note that the null momentum $P_u$ associated with $u$ takes the form $P_u = E - S$\,, with the total angular momentum playing the role of the charge $S$\,. 
There are several other reasons why this choice is relevant. 
Firstly, the centralizer of $P_u=E-S$ inside the $\mathfrak{so}(2,4)$ algebra spanned by the conformal Killing vectors of the metric \eqref{eq:lefdlc} composes the $\mathfrak{su}(1,2)$ subalgebra.
Secondly, this choice is reminiscent of other limiting procedures often considered in the literature, such as the Penrose and BMN limits \cite{Penrose1976,Berenstein:2002jq,Blau:2002dy}. 
Thirdly, it is natural in the context of SMT limits, which define a near-BPS sector of $\mathcal{N}=4$ SYM with SU(1\,,\,2) bosonic symmetry by requiring that the combination $E-S - Q \rightarrow 0$, where $Q$ is an appropriate sum over R-charges.\,\footnote{Since $Q$ commutes with all the generators of SU(2\,,\,2), this combination of R-charges is irrelevant when we deal with the SU(1\,,\,2) bosonic subgroup. 
Therefore, SU(1\,,\,2) arises as the centralizer of $E-S$ inside the SU(2\,,\,2) group.
In general, the SMT limit $E-S-Q \rightarrow 0$ preserves instead the super-group PSU$(1,2|3)$ \cite{Harmark:2007px,Harmark:2014mpa}. }
Lastly, the choice $P_u=E-S$ fits well with related limits for the string theory dual of SMT \cite{Harmark:2020vll}. 

Moreover, it is natural to take $E =  \, \p^{}_{x^0}$\,, such that the energy is conserved and is still associated with the new global time $x^0$ in the resulting non-Lorentzian CFT.\,\footnote{It is also possible to consider more complicated choices with $E + \alpha \, S =  \, \p^{}_{x^0}$ for $\alpha \neq -1$\,. For the purposes of building a state-operator correspondence in this work, the choice $\alpha=0$ will be sufficient.}
Combined with Eq.~\eqref{eq:nullv}, we find that the associated change of variables is given by
\beq \label{eq:tetc}
    \tau = R \, \bigl( x^0 - u \bigr)\,,
        \qquad%
    \eta = - 2 \, u \, .
\eeq
We therefore obtain the following null-reduction form of the line element~\eqref{eq:dsdsr}:
\beq 
\label{eq:nullredmetric}
    \dd s^2 = 2 \, \tau^{}_\mu \, \dd x^\mu \, \bigl( \dd u - m^{}_\nu \, \dd x^{\nu} \bigr) + h^{}_{\mu\nu} \, \dd x^\mu \, \dd x^\nu\,,
\eeq
where 
\begin{subequations} \label{eq:tnc}
\begin{align}
    \tau^{}_\mu \, \dd x^\mu & = R^2 \, \Bigl( \dd x^0 - \tfrac{1}{2} \, \cos \theta \, \dd \varphi \Bigr)\,,
        \qquad%
    h^{}_{\mu\nu} \, \dd x^\mu \, \dd x^\nu = \tfrac{1}{4} \, R^2 \, \bigl( \dd\theta^2 + \sin^2 \theta \, \dd\varphi^2 \bigr)\,, \\[4pt]
    m^{}_\mu \, \dd x^\mu & = \tfrac{1}{2} \, \Bigl( \dd x^0 + \tfrac{1}{2} \, \cos \theta \, \dd \varphi \Bigr)\,.
\end{align}
\end{subequations}
We have thus identified a mapping from the original coordinates $(\tau\,, \, \theta\,, \, \varphi\,, \, \eta)$ to the null-reduction coordinates $(x^0, \, \theta\,, \, \varphi\,, \, u)$\,, where $u$ is a \emph{null} isometry, in the sense that the $uu$ component of the metric vanishes and none of $\tau_\mu$\,, $m_\mu$ or $h_{\mu\nu}$ depends on $u$\,. 
Dimensionally reducing the $u$ direction gives rise to a TNC geometry described by the geometric data $\tau^{}_\mu$\,, $h^{}_{\mu\nu}$\,, and $m^{}_\mu$\,, with $\tau^{}_\mu$ the temporal vielbein, $h^{}_{\mu\nu}$ the spatial metric, and $m^{}_\mu$ the U(1) gauge potential associated with the local particle number symmetry. In the null-reduced nonrelativistic CFT, a fixed $P_u$ plays the role of a Bargmann mass, and a quadratic dispersion relation $E \sim k^2 / P_u + \cdots$ is developed, with $\mathbf{k}$ being the spatial momentum of a particle state. 
We note that the U$(1)$ gauge field $m_{\mu}$ includes a Dirac monopole $\tfrac{1}{2} \cos \theta \, \dd\varphi$ with a magnetic charge, which is expected to arise from dimensional reduction of $S^3$ in relation to the fact that $S^3$ can be described as a Hopf fibration over $S^2$. 

It is interesting to note that the TNC geometry~\eqref{eq:tnc} has a non-vanishing intrinsic torsion as $\dd \tau \neq 0$\,, which is in contrast to the Newton-Cartan geometry that covariantizes Newtonian gravity, since in the latter case $\dd \tau$ has to vanish identically. Furthermore, the TNC geometry~\eqref{eq:tnc} is neither free of the so-called \emph{twistless} torsion as $\tau \wedge \dd \tau \neq 0$\,, which implies that the temporal vielbein $\tau_\mu$ does \emph{not} sustain a well-defined foliation structure~\cite{Hartong:2022lsy}. This might lead to unwanted causal pathology if one associates $\tau_\mu \, \dd x^\mu$ with time. However, in the context of non-Lorentzian CFT defined by such a TNC, what we have learned is that $x^0$ (instead of $\tau$) is a globally well-defined time coordinate, as $\p_{x^0}$ corresponds to the conserved energy of the CFT. 
Note that, for fixed $(\theta\,, \, \varphi\,, \, u)$\,, the time evolution $\tau \rightarrow \tau + \dd \tau$ on the cylinder is indeed equivalent to $x^0 \rightarrow x^0 + \dd x^0$\,. It would be interesting to explore the consequence of the twist torsion in the future, which might be ultimately responsible to the causal structure that we discuss later in Section~\ref{sec:sph}.

\subsection{\texorpdfstring{$\mathfrak{su}$}{su}(1\,,\,2) Algebra}
\label{ssec:su12_algebra}

In the following, we show that the conformal Killing vector fields on the TNC geometry~\eqref{eq:tnc} form an $\mathfrak{su}(1,2)$ algebra, which further admits an extension to the $\mathfrak{su}(1\,,\,2)\oplus \mathfrak{u}(1)$ algebra. Geometrically, this central extension is realized by including the null circle in $u$\,. 

Note that the conformal Killing vectors of a four-dimensional Lorentzian manifold with coordinates $x^\text{M} = (x^\mu, \, u)$ are solutions to the equations: 
\begin{equation}
    \mathcal{L}^{\phantom{\dagger}}_{\Xi} \, g^{\phantom{\dagger}}_\text{MN} = 2 \, \omega \, g^{\phantom{\dagger}}_\text{MN} \, ,
\end{equation}
where $\mathcal{L}_{\Xi}$ denotes the Lie derivative along a conformal Killing vector $\Xi^\text{M} \, \p^{}_\text{M}$\,. In the null-reduced background \eqref{eq:nullredmetric}, the four-dimensional metric is decomposed into the clock one-form $\tau^{}_{\mu} = g^{}_{u \mu} \, \dd x^\mu$ and the spatial metric $h^{}_{\mu\nu}$ defined by $g^{}_{\mu\nu} = h^{}_{\mu\nu} - \tau^{}_\mu \, m^{}_\nu - \tau^{}_\nu \, m^{}_\mu$\,. The corresponding conformal Killing equations are then split into two sets of equations: 
\begin{align}\label{eq:RE-CKeq-Decom}
    \mathcal{L}_\xi \tau_\mu = 2 \, \omega \, \tau_\mu\,, 
        \qquad%
    \mathcal{L}_\xi h_{\mu\nu} + \tau_\mu \, \bigl(\mathcal{L}_\xi m_\nu - \partial_\nu \chi \bigr) + \tau_\nu \bigl( \mathcal{L}_\xi m_\mu -\partial_\mu \chi \bigr) = 2 \, \omega \, h_{\mu\nu}\,,
\end{align}
where we introduced the decomposition of the conformal Killing vector $\Xi^\text{M}=(\xi^\mu,\chi)$\,, with $\chi$ the $u$-component. 
We then define the Lie derivative with respect to the conformal Killing vector~$\xi^\mu \, \p_\mu$ according to the dilatation weights of the TNC geometric data as follows: 
\begin{align}\label{eq:TNC-CKequ}
\begin{split}
    \mathcal{L}_\xi \tau_\mu = 2 \, \omega \, \tau_\mu, 
        \qquad%
    \mathcal{L}_\xi h_{\mu\nu} = 2 \, \omega \, h_{\mu\nu}, 
        \qquad%
    \mathcal{L}_\xi m_\mu =  \partial_\mu \chi \,.
\end{split}
\end{align}
This is a natural conformal generalization of the scenario considered in~\cite{Grosvenor:2017dfs}, where it is required that the Lie derivative along a Killing vector on a non-Lorentzian manifold vanishes up to local Galilean transformations.\footnote{ 
Given the previous rules, the contravariant fields transform as $\mathcal{L}_\xi  v^\mu = -2 \, \omega \,v^\mu$ and $\mathcal{L}_\xi h^{\mu\nu} = -2 \, \omega \,h^{\mu\nu}$\,.
}
Solving Eq.~\eqref{eq:TNC-CKequ} for the TNC geometry~\eqref{eq:tnc}, we find the following conformal Killing vectors: 
\begin{subequations} \label{eq:su12-conformal-Killing}
\begin{align}
L_0 &= \frac{i}{4} \Bigl( 
3 \, \partial_{x^0} + \partial_u - 2 \, \partial_{\varphi}
\Bigr)\,, 
    \qquad%
\tilde{L}_0 = \frac{i}{4} \Bigl( 
3 \, \partial_{x^0} + \partial_u + 2 \, \partial_{\varphi}
\Bigr) \,, \\[4pt]
L_{\pm1} &= \frac{1}{4} \, e^{\mp i (x^0 -\frac{1}{2} \varphi)} \Bigl[ \pm \, 2 \cos  \tfrac{\theta}{2} \, \partial_{x^0} - 4 \, i \sin\tfrac{\theta}{2} \, \partial_{\theta} \pm \sec\tfrac{\theta}{2} \,
\bigl( \partial_{x^0} + \partial_u - 2 \, \partial_{\varphi} \bigr) 
\Bigr] , \\[4pt]
\tilde{L}_{\pm1} &= \frac{1}{4} e^{\mp i (x^0 +\frac{1}{2} \varphi)} \Bigl[ 
\mp \, 2 \sin \tfrac{\theta}{2} \, \partial_{x^0} - 4 \, i \cos\tfrac{\theta}{2} \, \partial_{\theta} \mp \csc\tfrac{\theta}{2} \, 
\bigl( \partial_{x^0} + \partial_u + 2 \, \partial_{\varphi} \bigr)
\Bigr] , \\[4pt]
J_\pm & = i \, e^{\pm i \varphi} \left[ 
\pm \, i \, \partial_\theta - \cot \theta \, \partial_\varphi - \frac{1}{2\sin \theta} \, \bigl( \partial_{x^0} +\partial_u \bigr)
\right].
\end{align}
\end{subequations}
This set of generators forms an $\mathfrak{su}(1\,,2)$ algebra, 
with non-vanishing commutators given by
\begin{subequations} \label{eq:suot}
\begin{align} 
	[L_m\,,\,L_n] &= (n-m) \, L_{m+n}, 
        &%
    [L_0\,,\,\tL_n] &= -\tfrac{n}{2} \, \tilde{L}_n, 
        &%
    \bigl[ L_{\pm1}\,, \, J_\mp \bigr] &= \mp \tilde{L}_{\pm1} \\[4pt]
    [\tilde{L}_{m}\,,\,\tL_n] &= (m-n) \, \tL_{m+n}\,,
        &%
    [\tL_0\,,\,L_n] &= \tfrac{n}{2} \, L_n\,, 
        &%
    \bigl[ \tilde{L}_{\pm1}\,, J_{\pm} \bigr] &= \mp L_{\pm1}\,, \\[4pt]
    \bigl[ L_0\,,\,J_\pm \bigr] &= \pm \tfrac{1}{2} J_\pm\,,  
        &%
    \bigl[ L_{\pm1}\,,\,\tilde{L}_{\mp1} \bigr] &= \mp J_\pm\,,
        &%
    \bigl[ J_+\,, \, J_- \bigr] & = 2 \, \bigl( L_0 - \tilde{L}_0 \bigr)\,,
    \\[4pt]
    \bigl[ \tilde{L}_0\,,\,J_\pm\bigr] &= \mp \tfrac{1}{2} J_\pm\,, 
\end{align}
\end{subequations}
where $m\,,\,n = -1\,,\, 0\,,\, 1$\,. 
This algebra is composed by two coupled $\mathfrak{su}(1,1)$ algebra, which are spanned by the set of $L_{m} (\tilde{L}_{m})$ generators, respectively. 
Note that the generators in the algebra commute with the null translation generator $\partial_u$\,. The complete underlying algebra is then $\mathfrak{su}(1,2) \oplus \mathfrak{u}(1)$\,, which also contains the additional $\mathfrak{u}(1)$ central extension associated with the generator $\partial_u$\,. We will discuss this central extension further in Section~\ref{sec:state_ope_map}. 
Note that the TNC conformal Killing equations~\eqref{eq:TNC-CKequ} are only sufficient but not necessary conditions to solve the relativistic conformal Killing equations~\eqref{eq:RE-CKeq-Decom} in the parent theory. 
The solutions to Eq.~\eqref{eq:TNC-CKequ} therefore only span a subset of the conformal Killing vectors on the original Lorentzian manifold, where the generators form an $\mathfrak{so}(2,4)$ algebra. 
This $\mathfrak{su}(1,2)$
subset of the $\mathfrak{so}(2,4)$ algebra is spanned by the closed set of generators that commute with the translation along the null direction. 

In Appendix~\ref{app:su12_algebra}, we present further details of the $\mathfrak{su}(1,2) \oplus \mathfrak{u}(1)$ algebra. We summarize a few salient properties here. First, 
there exists a contraction of this algebra that leads to 
the three-dimensional Schr\"{o}dinger algebra (see Appendix~\ref{app:su12_algebra} for the precise matching). Second, the three-dimensional Schr\"{o}dinger algebra admits a triple extension, which was found as a conformal completion of the $z = 2$ Lifshitz algebra in~\cite{Hartong:2016yrf}.
It was later understood that this extended Schr\"odinger algebra can be obtained from the leading-order terms of an expansion of the $\mathfrak{su}(1\,,\,2)\oplus \mathfrak{u}(1)$ algebra at large speed of light $c \to \infty$ \cite{Kasikci:2020qsj}.%
\footnote{The expansion parameter $c$ was not interpreted as speed of light in \cite{Kasikci:2020qsj}. 
But this algorithm manifestly fits into the large speed of light expansion algorithm in \cite{Hansen:2020pqs,Hansen:2019vqf}. 
} 
The explicit map is reported in Eq.~\eqref{eq:mapsbetweenalgebra}. Henceforth, we will take the $\mathrm{SU}(1\,,2) \times \mathrm{U}(1)$ symmetry as the non-Lorentzian conformal group that underlies our analysis of the non-Lorentzian state-operator correspondence.

\subsection{Operator Picture from an Infinite Radius Limit of State Picture}
\label{ssec:infinite_radius_state}

We are ready to perform the analogous analysis as for the relativistic case in Section~\ref{sec:lt}, but now for non-Lorentzian SU(1\,,\,2) CFTs. We currently have in our hand the TNC geometry~\eqref{eq:tnc} in the state picture. Next, we would like to derive the targeting geometry in the operator picture, which will then also facilitate us with the explicit construction of the coordinate transformation mapping between these two geometries. 

We recall that, in relativistic CFTs, there is a simple relation between the Minkowski spacetime~\eqref{mink_sphe} and the Lorentzian cylinder~\eqref{eq:conmappedMink}: the Minkowski spacetime arises from sending the radius of the Lorentzian cylinder to infinity. 
We now implement this relation explicitly, starting with the four-dimensional Lorentzian cylinder~\eqref{eq:lefdlc}, where we introduce the coordinate change
\beq \label{eq:ttve}
    t = \tau\,,
        \qquad%
    \theta = \frac{\pi}{2} - \frac{2 \, y}{R}\,,
        \qquad%
    \varphi = \frac{2 \, x}{R}\,,
        \qquad%
    \eta = \frac{2 \, z}{R}\,.
\eeq
Plugging Eq.~\eqref{eq:ttve} into the line element~\eqref{eq:lefdlc}, followed by an expansion with respect to large $R$\,, we find
\beq
\label{eq:Mink_metric}
    \dd s^2 = - \dd \tau^2 + \dd x^2 + \dd y^2 + \dd z^2 + O\bigl( R^{-1} \bigr)\,.
\eeq
Indeed, in the infinite $R$ limit, the Minkowski spacetime in the operator picture is recovered. Furthermore, combining Eqs.~\eqref{eq:ttve} and~\eqref{eq:tetc}, we find
\beq \label{eq:wm}
    x^0 = \frac{\tau - z}{R}\,,
        \qquad%
    u = - \frac{z}{R}\,.
\eeq
Of course, the $R \rightarrow \infty$ limit of the null-reduction form~\eqref{eq:nullredmetric} still gives the same Minkowski metric~\eqref{eq:Mink_metric}. We simply do not have access to the non-Lorentzian geometry in the operator picture that corresponds to Eq.~\eqref{eq:tnc}.  

Clearly, the above prescription needs to be modified in nonrelativistic CFTs. The reason that the na\"{i}ve mapping~\eqref{eq:wm} does not work is because it does not properly take into account the null circle in $u$\,, while the desired map should be between the TNC geometry~\eqref{eq:tnc} in the state picture and some other non-Lorentzian geometry in the operator picture, which both supposedly arise after the same null reduction is performed. The modification to Eq.~\eqref{eq:wm} is simple: we should perform the transformations intrinsically at the level of the TNC geometry~\eqref{eq:tnc}. In practice, we define the following new coordinate change:
\beq \label{eq:nlirl}
    x^0 = \frac{T}{R^2}\,,
        \qquad%
    \theta = \frac{\pi}{2} - \frac{2 \, y}{R}\,,
        \qquad%
    \varphi = \frac{2 \, x}{R}\,,
\eeq
Note that $T$ will be the new operator-picture time and we have introduced a further rescaling of $x^0$ such that the TNC data in Eq.~\eqref{eq:tnc} are non-singular in the $R \rightarrow \infty$ limit. 
Finally, plugging Eq.~\eqref{eq:nlirl} into the TNC geometry~\eqref{eq:tnc}, we find the appropriate geometry underlying the operator picture in the infinite $R$ limit,
\beq \label{eq:nlgop}
    \hat{\tau}^{}_\mu \, \dd x^\mu = \dd T - 2 \, y \, \dd x\,,
        \qquad%
    \hat{h}^{}_{\mu\nu} \, \dd x^\mu \, \dd x^\nu = \dd x^2 + \dd y^2\,,
        \qquad%
    \hat{m}^{}_\mu \, \dd x^\mu = 0\,.
\eeq
Furthermore, redefining the operator-picture time as $T \rightarrow T + x \, y$\,, we bring the geometric data in Eq.~\eqref{eq:nlgop} to the form
\beq \label{eq:mnlgop}
    \hat{\tau}^{}_\mu \, \dd x^\mu = \dd T + \bigl( x \, \dd y - y \, \dd x \bigr)\,,
        \qquad%
    \hat{h}^{}_{\mu\nu} \, \dd x^\mu \, \dd x^\nu = \dd x^2 + \dd y^2\,,
        \qquad%
    \hat{m}^{}_\mu \, \dd x^\mu = 0\,.
\eeq
This is the TNC geometry corresponding to the operator-picture counterpart of Eq.~\eqref{eq:tnc}, which was derived in the state picture. This geometry
fits in a set of backgrounds admitting an $\Omega$-deformation which is crucial for the realization of a corresponding SU(1\,,\,$n$) conformal isometry (for any $n>1$) \cite{Lambert:2021nol}.\,\footnote{One way to introduce an $\Omega$-deformation in a non-Lorentzian geometry is to perform a null reduction of Minkowski space followed by a Weyl rescaling \cite{Lambert:2021nol}. We outline this procedure in Appendix~\ref{app:ssec:coord_ope}, but using different conventions compared to \cite{Lambert:2021nol}, such that now the $\Omega$-deformation is dimensionless.}
In the present case, the $\Omega$-deformation reads $x \, \dd y - y \, \dd x$\,, and we verified indeed that the associated conformal Killing vector fields generate the SU(1\,,\,2) conformal group, see Appendix~\ref{app:su12_algebra}. 
Furthermore, it is interesting to notice that the geometry~\eqref{eq:mnlgop} respects a scaling symmetry of the form
\begin{equation}
    u \rightarrow u\,, 
        \qquad%
    T \rightarrow b^2 \, T\,, 
        \qquad%
    x^i \rightarrow b \, x^i\,,
\label{eq:z2_Lifhsitz_scaling}
\end{equation}
where $b$ is a constant.
This scaling with dynamical exponent $z=2$ is reminiscent of a direct null reduction procedure in Minkowski space, and it is consistent with the following engineering scaling dimensions of the coordinates:
\beq
[u] = 0 \, , \qquad
[T] = 2 \, , \qquad
[x^i] = 1 \, .
\label{eq:engineer_dim}
\eeq
In Eq.~\eqref{eq:scaling_Minkowski_Lambert}, the same scaling dimensions arise as a consequence of the coordinate transformation~\eqref{eq:coord_transf_Lambert_ourconv}.

Finally, we define the polar coordinates for the $\Omega$-deformed TNC geometry~\eqref{eq:mnlgop}. 
A particularly useful coordinate change is 
\beq 
    x = 2 \, \hat{r} \, \cos \hat{\varphi}\,,
        \qquad%
    y = 2 \, \hat{r} \, \sin \hat{\varphi}\,,
        \qquad%
     T = \frac{\hat{t}}{2} - \hat{r}^2 \, \sin \bigl( 2 \, \hat{\varphi} \bigr)\,,
\eeq
such that the geometric data take the following relatively simple form:
\beq \label{eq:hthm}
    \hat{\tau}^{}_\mu \, \dd x^\mu = 2 \, \dd \hat{t} + 4 \, \hat{r}^2 \, \dd\hat{\varphi}\,,
        \qquad%
    \hat{h}^{}_{\mu\nu} \, \dd x^\mu \, \dd x^\nu = 4 \, \bigl( \dd \hat{r}^2 + \hat{r}^2 \, \dd \hat{\varphi}^2 \bigr)\,,
        \qquad%
    \hat{m}^{}_\mu \, \dd x^\mu = 0\,.
\eeq
This is the parametrization of the TNC geometry in the operator picture that we will use below. 
In particular, it can be directly obtained from an $R \rightarrow \infty$ limit of the state-picture background \eqref{eq:tnc}, as we will comment around Eq.~\eqref{eq:map_leading_order} in Section~\ref{ssec:exact_map}.

\subsection{Conformal Mapping between Torsional Newton-Cartan Geometries}
\label{ssec:conf_map_NCgeo}

Now that we have obtained the TNC geometries on both sides of the non-Lorentzian state-operator correspondence, it is a conceptually straightforward but yet technically challenging problem to construct the explicit mapping. We will first present the result below and then give the detailed derivation using the technique of null reduction later in Section~\ref{sec:state_ope_map}. 

Notice that in order to build a map between the two geometries at finite $R$, we need to map the angular coordinate $\varphi$ on the state side to a $\mathrm{U}(1)$ direction on the operator side. For this reason, we consider the change of variables from the state-picture polar coordinates $(x^0\,, \, \theta\,, \, \varphi)$ to the operator-picture polar coordinates $(\hat{t}\,, \, \hat{r}\,, \, \hat{\varphi})$, rather than the Cartesian ones. 
The map reads
\begin{subequations} \label{eq:cmnr}
\begin{align}
    x^0 + \tfrac{1}{2} \, \varphi & = - \mathrm{arccot} \! \le \frac{\mathbf{t}^2 + \mathbf{r}^4 - 1}{\mathbf{t}} \ri,  \\[4pt]
    \varphi & = \hat{\varphi} - \mathrm{arctan} \! \le \frac{\mathbf{r}^2 - 1}{\mathbf{t}} \ri,  \\[4pt] 
    \theta &=  \mathrm{arccos} \! \left[ 1 - \frac{8 \, \mathbf{r}^2}{\mathbf{t}^2 + \bigl( \mathbf{r}^2 + 1 \bigr)^2} \right], \label{eq:phitheta}
\end{align}
\end{subequations}
where we have performed a $z=2$ rescaling to define 
\bea
    \mathbf{t} \equiv \frac{\hat{t}}{2 \, R^2}\,,
        \qquad%
    \mathbf{r} \equiv \frac{\hat{r}}{2 \, R}\,.
    \label{eq:bold_tr}
\eea
This change of variables maps the TNC data $(\tau_{\mu}\,, \, m_{\mu}\,, \, h_{\mu\nu})$ to $(\hat{\tau}_{\mu}\,, \, \hat{m}_{\mu}\,, \, \hat{h}_{\mu\nu})$ up to the three-dimensional Weyl rescaling, local Galilean boost, and U(1) gauge transformation. 

Nevertheless, the three-dimensional coordinate transformations in Eq.~\eqref{eq:cmnr} are not sufficient to capture the complete mapping from the state to operator picture. This can be made manifest by resorting to the null reduction of the four-dimensional relativistic system, where the mapping between the null directions in the state and operator pictures is non-trivial and takes the following form:
\bea \label{eq:uuhattrnsf}
    u = \hat{u} + \frac{1}{2} \left[ \hat{\varphi} + \, \mathrm{arctan} \! \le \frac{\mathbf{r}^2 - 1}{\mathbf{t}} \ri \right].
\eea
Using the first equation in Eq.~\eqref{eq:phitheta}, we find that the transformation~\eqref{eq:uuhattrnsf} can be equivalently written as
\bea
    u + \tfrac{1}{2} \, \varphi = \hat{u} + \hat{\varphi}\,. 
    \label{eq:uuhat2}
\eea
In the null-reduced theory, the generator corresponding to the translation along the null direction is associated with a U(1) gauge field. This translation generator in the extra dimension acquires the interpretation of the particle number generator in the three-dimensional nonrelativistic CFT. Although the relation between $u$ and $\hat{u}$ is nontrivial, the three-dimensional coordinate transformation~\eqref{eq:cmnr} is independent of $u$\,. This guarantees that the generator for the U(1) gauge symmetry is preserved in both pictures, \ie $\p^{}_u = \p^{}_{\hat{u}}$\,, which implies that the null reductions are equivalent. 
However, the transformation~\eqref{eq:uuhattrnsf} does impact the mapping between the Hamiltonians, which we discuss in the next subsection.

\subsection{Mapping the Generators}
\label{ssec:map_generators}

Having found an explicit map between the two TNC geometries~\eqref{eq:tnc} and~\eqref{eq:hthm} with SU(1\,,\,2) conformal isometry, we are finally ready to describe a realization of the nonrelativistic state-operator correspondence.
To achieve this task, we first need to understand the relation between the generators on both sides, which are responsible for either the construction of the Hilbert space, or the insertion of local operators with definite scaling dimension.

\subsubsection{State-Picture Hamiltonian} \label{sec:sph}

In the state picture, the Hilbert space is built out of the eigenstates of the Hamiltonian 
\bea
    H_0 = \p_{x^0}\,,
\eea
which generates the time translation between equal-time slices in the state picture described by the geometry~\eqref{eq:tnc}. This Hamiltonian is analogous to $H^{\rm L}$ on the Lorentzian cylinder in the relativistic case, which we have described in Section~\ref{sec:lt}.
In order to realize a state-operator correspondence, we rewrite the Hamiltonian $H_0$ in terms of the generators of the $\mathfrak{su}(1,2) \oplus \mathfrak{u}(1)$ algebra in the operator picture, and then relate the latter with the action of inserting local operators with definite scaling dimension at a specific point.

We begin by discussing what the state-picture Hamiltonian $H_0$ corresponds to in the operator picture. This is achieved by applying the coordinate transformation~\eqref{eq:cmnr} and the shift of the extra null coordinate in Eq.~\eqref{eq:uuhattrnsf} to $H_0$\,, which gives
\beq
H_0 = \frac{1}{2} \Bigl[  - \le \mathbf{r}^2 + 1 \ri  \, \p^{}_{\hat{u}}
+ \le \mathbf{t}^2 - \mathbf{r}^4 +1 \ri \, \p_{\mathbf{t}}
+ \mathbf{t} \, \mathbf{r} \, \p_{\mathbf{r}} 
+  \le \mathbf{r}^2 + 1 \ri \p_{\hat{\varphi}} \Bigr] \, .
\label{eq:diff_rep_H0_ope}
\eeq
Here, the generator $N \equiv \p^{}_{\hat{u}}$ appears in the operator picture because the null circles in the parent four-dimensional relativistic theory are non-trivially related as in Eq.~\eqref{eq:uuhat2}. From the perspective of the nonrelativistic CFT obtained from the null reduction along the compact $\hat{u}$ direction, the eigenvalue associated with $\p_{\hat{u}}$ corresponds to a central charge, which forms the U(1) sector of the SU(1\,,\,2)$\times$U(1) symmetry. 
In the Schr\"{o}dinger basis, this eigenvalue corresponds to the mass of the nonrelativistic particle, and the generator $N$ corresponds to the particle number conservation. 

\begin{figure}[ht]
    \centering
    \hspace{9mm}\includegraphics[width=0.9\linewidth]{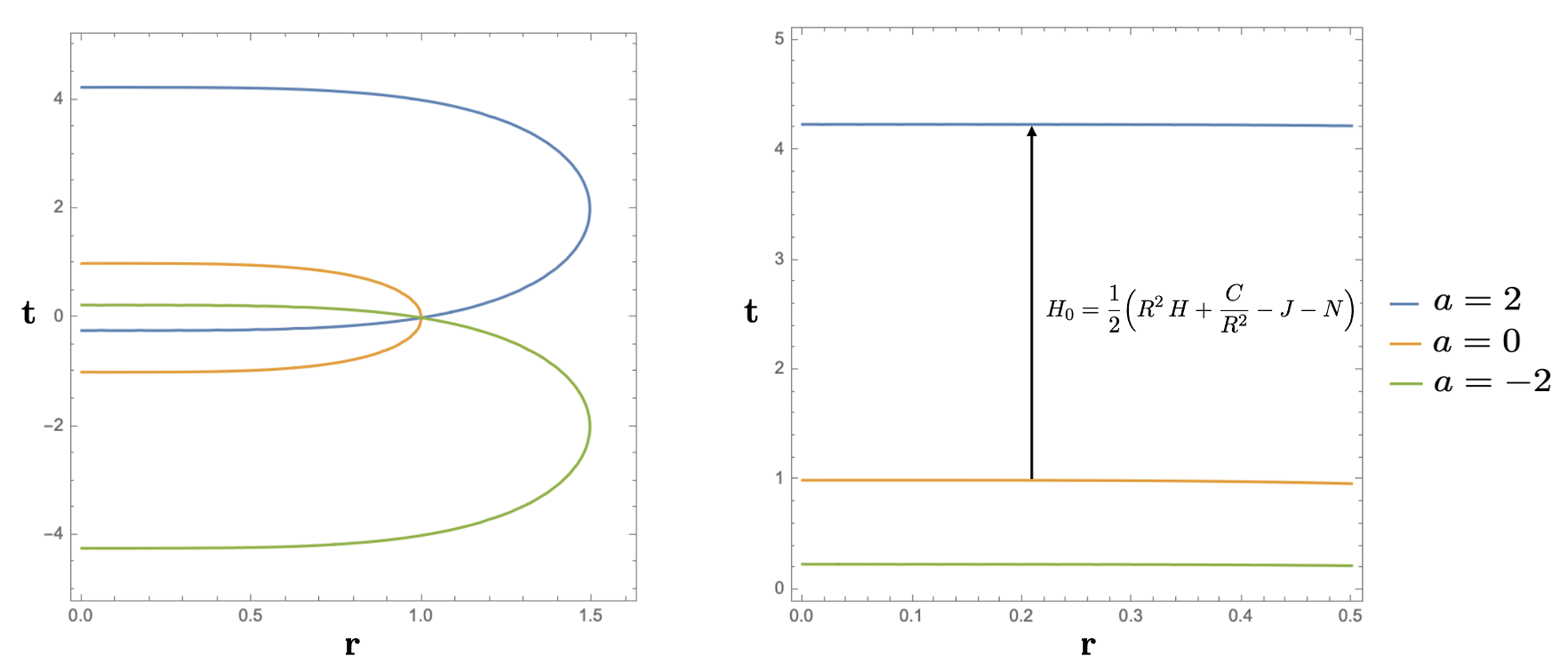}
    \qquad
    \caption{\textbf{Left:} Plot of the curves \eqref{eq:quartic_curves_su12} in the $(\mathbf{t}\,,\mathbf{r})$ plane. \textbf{Right:} At small $\mathbf{r}$ with $\mathbf{r} < 1$\,, there is \emph{no} intersection between the equal-time hypersurfaces, and the spacetime causality is well defined.     }
    \label{fig:curves_su12}
\end{figure}

In the operator picture, the differential representations of the Hamiltonian generator $H$\,, angular momentum generator $J$\,, and special conformal transformation generator $C$ are
\begin{align}
    H = \frac{\p^{}_{\mathbf{t}}}{R^2}\,,
        \qquad%
    J = - \p^{}_{\hat{\varphi}}\,,
        \qquad%
    C = R^2 \, \Bigl[ - \mathbf{r}^2 \, \p^{}_{\hat{u}} + \left( \mathbf{t}^2 - \mathbf{r}^4 \right) \p^{}_{\mathbf{t}} + \mathbf{t} \, \mathbf{r} \, \p^{}_{\mathbf{r}} + \mathbf{r}^2 \, \p^{}_{\hat{\varphi}} \Bigr]\,.
    \label{eq:ope_generators}
\end{align}
Combined with Eq.~\eqref{eq:diff_rep_H0_ope}, we find
\beq
    H_0 = \frac{1}{2} \Bigl( R^2 \, H + \frac{C}{R^2} - J - N \Bigr) \,.
    \label{eq:H0_ope}
\eeq
Here, the generator $N$ is associated with the central charge $m$\,, which is interpreted as the Bargmann mass of the nonrelativistic particle. 
The existence of a non-trivial $m$ in the generator is a consequence of non-trivial gauge transformations that needs to be performed in the TNC geometry to map the state and operator backgrounds.
Note that the constant $x^0$ slices are now mapped in the operator picture via the coordinate transformation \eqref{eq:cmnr} to the surfaces\,\footnote{They can be either be obtained as the integral curves generated by the differential operator \eqref{eq:diff_rep_H0_ope}, or by directly looking at the transformation of $x^0$ at fixed $\varphi$ in Eq.~\eqref{eq:cmnr}.}
\beq
(\mathbf{t}-a)^2 + \mathbf{r}^4 = 1+a^2  \, ,
\label{eq:quartic_curves_su12}
\eeq
where $a=-\cot x^0$ is an integration constant at fixed $x^0$ and $\varphi$.
These surfaces are sketched in Fig.~\ref{fig:curves_su12} for a selection of $a$'s.
Comparing the previous expression with the hyperbolae in Eq.~\eqref{eq:hyperbolae_rel_case} arising in the relativistic case, there are two major differences.
Firstly, the defining equation is quartic, reflecting the $z=2$ Lifshitz scaling of the time and space coordinates in the nonrelativistic case.
Secondly, the curves are closed and do \emph{not} provide a foliation of the geometry, which is reminiscent of the existence of the twist torsion in the TNC geometry that we have discussed at the end of Section~\ref{sec:nullred_RS3}. In contrast, when $\mathbf{r} < 1$\,, we regain the notion of an absolute time direction and hence a well-defined causal structure.

\subsubsection{Construction of the Hilbert Space} 

Finally, we discuss the realization of the state-operator correspondence in non-Lorentzian CFTs. This realization is two-fold: First, one has to construct the Hilbert space by acting with local operators on the vacuum $\ket{0}$\,.\,\footnote{
Here, local operators act at the origin of the spacetime in the operator picture. This is not restrictive, since one can reach an arbitrary point $(\mathbf{t}\,,\,\mathbf{r})$ starting from $(0,0)$ by applying the generators in SU$(1,2)$. }
Second, one has to build a one-to-one correspondence between the eigenvalues of $H_0$ in Eq.~\eqref{eq:H0_ope} and the scaling dimensions of the associated local operators. 
Before we discuss the SU(1\,,\,2) case, we first review the algebraic arguments in~\cite{Nishida:2007pj} for Schr\"{o}dinger-invariant CFTs and the extension to SU(1\,,\,$n$)-invariant CFTs outlined in~\cite{Lambert:2021nol}.
We will then complement and justify similar arguments in the SU(1\,,\,2) case using the map between state and operator geometries that we have found in Section~\ref{ssec:conf_map_NCgeo}.

\vspace{3mm}

\noindent $\bullet$~\emph{Schr\"{o}dinger conformal field theories}

\vspace{1mm}

--- \emph{operator picture} ---

\vspace{1mm}

We begin with the Schr\"{o}dinger-invariant CFTs~\cite{Nishida:2007pj} in the operator picture.
Local operators $\mathcal{O}\bigl(\hat{t}\,,\,\hat{x}^i\bigr)$ in the Heisenberg picture are defined at an arbitrary point on a manifold in terms of operators inserted at the origin $\mathcal{O}(0)$ as\,\footnote{Without loss of generality, the origin in the operator picture is defined by $\hat{t}=\hat{r}=0$\,.} 
\beq
\mathcal{O}(\hat{t}\,,\,\hat{x}^i) = e^{i \, (H \, \hat{t} - P_i \, \hat{x}^i)} \, \mathcal{O}(0) \, 
e^{-i \, (H \, \hat{t} - P_i \, \hat{x}^i)} \, ,
\label{eq:local_operator_space}
\eeq
where $H$ and $P_i$ are the Hamiltonian and momentum generators in the operator picture, respectively.
We consider the stability group that leaves the origin of the geometry invariant. This stability group is spanned by the generators associated with the spatial rotation~$J$\,, dilatation~$D$\,, special conformal transformation~$C$\,, and the~$\mathrm{U}(1)$ central extension~$N$\,.
The local operators at the origin form a representation of this reduced group. The Schr\"{o}dinger algebra formed by the above generators is
\begin{subequations}\label{eq:sa}
\begin{align}
    [H\,,\,G_i] &= P_i\,, 
        &%
    [P_i\,,\,C] &= G_i\,, 
        &%
    [H\,,\,C] &= D, 
        &%
    [P_i\,,\,G_j] &= \delta_{ij}\, N, \\[4pt]
    [D\,,\,H] &= -2 \, H,
        &%
    [D\,,\,C] &= 2 \, C, 
        &%
    [D\,,\,P_i] &= -P_i\,,
        &%
    [D\,,\,G_i] &= G_i\,, \\[4pt]
    [J\,,\,P_i] &=\epsilon_{ij} \, P_j\,, 
        &%
    [J\,,\,G_i] &=\epsilon_{ij} \, G_j\,.
\end{align}
\end{subequations}
For a local operator $\mathcal{O}$ in the Schr\"{o}dinger CFT, we define the scaling dimension $\Delta$ and the eigenvalue $m$ associated with the central $\mathrm{U}(1)$ generator $N$ to be
\beq
    \bigl[ D\,,\, \mathcal{O}(0) \bigr] =  \Delta \,  \mathcal{O}(0)\,,  
        \qquad%
    \bigl[ N\,,\, \mathcal{O}(0) \bigr] = m \, \mathcal{O}(0)\,.
\label{eq:eigenvalues_ope_Son}
\eeq
The commutation relations of the Schr\"{o}dinger algebra \eqref{eq:sa} identify $H$ and $P_i$ as the raising operators, and $C$ and $G_i$ as the lowering operators. We then identify the primary operators that have the lowest weights in the representation by requiring that they vanish under the action of the lowering operators, \emph{i.e.},
\beq
    \bigl[ G_i\,,\, \mathcal{O}(0) \bigr] = \bigl[ C\,, \, \mathcal{O}(0) \bigr] = 0 \, .
\label{eq:primary_ope}
\eeq
The full tower of descendants can then be obtained by acting $H$'s and $P_i$'s on the primaries. 

\vspace{1mm}

--- \emph{state picture} --- 

\vspace{1mm}

Next, we build the state-operator correspondence and thus pass on to the state picture. Consider a primary operator $\mathcal{O}(0)$ inserted at the origin of the operator picture. The corresponding state is given by
\beq \label{eq:state_Son}
    \ket{\Psi_{\mathcal{O}}} \equiv e^{-H} \, \mathcal{O}^{\dagger}(0) \, \ket{0}\,.
\eeq
We assume that $\mathcal{O}^\dagger$ consists of creation operators. 
It is shown in Eq.~(46) of~\cite{Nishida:2007pj} that~$\ket{\Psi_{\mathcal{O}}}$ is an energy eigenstate of an oscillator Hamiltonian~$H_{\rm osc}$ with eigenvalues given by the scaling dimension of the local operator:\,\footnote{The quantity $H_{\rm osc}$ is called oscillator Hamiltonian because the generator $C$ is quadratic in position space.}
\beq
H_{\rm osc} \ket{\Psi_{\mathcal{O}}}= \Delta \ket{ \Psi_{\mathcal{O}}} \, , \qquad
H_{\rm osc} = H+C \, .
\label{eq:Hosc}
\eeq
Moreover, there exist ladder operators $\ell_{\pm} = H_{\rm osc} \pm D $ such that~$\ket{\Psi_{\mathcal{O}}}$ is the lowest weight, \ie
\beq
\ell_- \ket{\Psi_{\mathcal{O}}} = 0 \, .
\eeq
and the full ladder of descendant states are built by acting with $\ell_+$'s.
This concludes the realization of the state-operator correspondence in the Schr\"{o}dinger case.
The validity of this construction has been tested in several physical settings, where the spectrum of the dilatation operator was found by instead computing the energy spectrum of particles confined in a harmonic potential \cite{Nishida:2007pj,Nishida:2010tm}.

\vspace{3mm}
\noindent $\bullet$~\emph{SU(1\,,\,2)$\times$U(1) conformal field theories.} It is shown in~\cite{Lambert:2021nol} that, in CFTs with the $\mathfrak{su}(1\,,n)$ ($n \geq 1$) symmetry, the Hilbert space can be built analogouly as in Schr\"{o}dinger CFTs. Here, we focus on the case of interest with $n = 2$\,, where the algebra is given in Eq.~\eqref{eq:suot}. Consider the following automorphism:
\begin{subequations}\label{eq:automorphismsu12}
\begin{align}
    L_0 &= \frac{1}{4} \, \bigl( D + 3 \, i \, J \bigr)\,,
        &%
    L_1 &= - \frac{i}{\ell} \, G\,, 
        &%
    L_{-1} &= \ell \, P^*\,, \\[4pt]
    \tilde{L}_0 &= - \frac{1}{4} \, \bigl( D - 3 \, i \, J \bigr)\,, 
        &%
    \tilde{L}_1 &= \tilde{\ell} \, P\,,
        &%
    \tilde{L}_{-1} &= \frac{i}{\tilde{\ell}} \, G^*\,, \\[4pt]
        &
        &%
    J_+ &= \frac{i \, C}{\ell \, \tilde{\ell}}\,,
        &%
    J_- &= i \, \ell \, \tilde{\ell} \, H\,,
\end{align}
\end{subequations}
for arbitrary constants $\ell$ and $\tilde{\ell}$\,. Here, 
\bea
    P = \frac{1}{\sqrt{2}} \, \bigl( P_1 + i \, P_2 \bigr)\,,
        \qquad%
    G = \frac{1}{\sqrt{2}} \, \bigl( G_1 + i \, G_2 \bigr)\,. 
\eea
The $\mathfrak{su}(1\,,2)$ algebra~\eqref{eq:suot} becomes in form the same as the Schr\"{o}dinger algebra~\eqref{eq:sa} except that the following commutators are modified:
\begin{align} \label{eq:msuot}
    \bigl[ P_i\,, \, G_j \bigr] = - \tfrac{3}{2} \, \delta_{ij} \, J + \tfrac{1}{2} \, \epsilon_{ij} \, D\,,
        \qquad%
    \bigl[ P_i\,,\, P_j \bigr] = \epsilon_{ij} \, H\,,
        \qquad%
    \bigl[ G_i\,,\, G_j \bigr] = \epsilon_{ij}\, C\,.
\end{align}
The $\mathfrak{u}(1)$ extension shifts $J$ in the commutator $[P_i\,, \, G_j]$ in Eq.~\eqref{eq:msuot} by the central charge generator $N$\,, such that 
\bea \label{eq:msuot2}
    \bigl[ P_i\,, \, G_j \bigr] = - \tfrac{3}{2} \, \delta_{ij} \, \bigl( J + N \bigr) + \tfrac{1}{2} \, \epsilon_{ij} \, D\,.
\eea
We are thus led to the $\mathfrak{su}(1\,,2) \oplus \mathfrak{u}(1)$ that underlies our non-Lorentzian CFT. Note that the Schr\"{o}dinger algebra~\eqref{eq:sa} can be viewed as a contraction of the $\mathfrak{su}(1\,,2) \oplus \mathfrak{u}(1)$ algebra by introducing the rescaling
\bea \label{eq:rescaling-PG}
    P_i \rightarrow   P_i/\varepsilon\,,
        \qquad%
    G_i \rightarrow  G_i/\varepsilon \,,
\eea
followed by sending $\varepsilon$ to zero. 

The construction of the Hilbert space then follows almost identically as we have discussed in Schr\"{o}dinger CFTs, which we briefly recap below. In the operator picture, among the $\mathfrak{su}(1\,,2) \oplus \mathfrak{u}(1)$ generators, we again identify $H$ and $P_i$ as the raising operators, and $C$ and $G_i$ the lowering operators. One may then apply these raising and lowering operators to build the Hilbert space. The primary operators and states are defined in the same way as in Schr"odinger CFTs, except that now the oscillator Hamiltonian~\eqref{eq:Hosc} is replaced with the Hamiltonian that we constructed earlier in Eq.~\eqref{eq:H0_ope}. Using Eq.~\eqref{eq:eigenvalues_ope_Son}, the state-picture Hamiltonian $H_0$ is now represented in terms of the operator-picture quantities as
\bea
    H_0 = R^2 \, H + \frac{C}{R^2} - J - m\,,
    \label{eq:state_pic_HamH0}
\eea
while the ladder operators are $\ell_\pm = R^2 H + \frac{C}{R^2} \pm D$\,.

It is interesting to compare our results with the relevant discussions in~\cite{Lambert:2021nol}, where a similarity transformation is performed on the space of operators. Denote the similarity transformation of a generator $X$ as $\bar{X}$\,. In particular, among the transformed operators, we have
\beq
    \bar{D} = -i \le \mu \, H + \frac{C}{\mu} \ri,
\label{eq:Dbar_Lambert}
\eeq
where $\mu$ is a length scale. 
Note that the map~\eqref{eq:Dbar_Lambert} in~\cite{Lambert:2021nol} assumes that the time and space coordinates have the same length dimension. In our conventions~\eqref{eq:engineer_dim}, we identify $\mu \rightarrow R^2$\,. 
By analogy, there exists an automorphism that leads to the identity~\eqref{eq:rel_automorphism} in the Lorentzian case \cite{Minwalla:1997ka}.
Note that primary states $\ket{\Psi_{\mathcal{O}}}$ of the form \eqref{eq:state_Son} satisfy~\cite{Lambert:2021nol}
\begin{subequations}
\begin{align}
    \bar{D} \ket{\Psi_{\mathcal{O}}} &= \Delta \ket{\Psi_{\mathcal{O}}} \, , 
        &%
    \bar{N} \ket{\Psi_{\mathcal{O}}} &= m \ket{\Psi_{\mathcal{O}}}, 
        &%
    \bar{J} \ket{\Psi_{\mathcal{O}}} &= j \ket{\Psi_{\mathcal{O}}}, \\[4pt]
        &&%
    \bar{G}_i \ket{\Psi_{\mathcal{O}}} &= 0 \, , 
        &%
    \bar{C} \ket{\Psi_{\mathcal{O}}} &= 0 \, , 
\end{align}
\end{subequations}
while the action of $\bar{H}$ and $\bar{P}_i$ on the primary states creates towers of descendants.
It then follows that states in the Hilbert space are eigenvectors of the generator $\bar{D}$. Their eigenvalues are precisely the scaling dimensions of the operators creating the states.
This way of using an automorphism (up to an analytic continuation) of the generators to build an operator $\bar{D}$, whose eigenstates provide an irreducible representation of the algebra, was also considered previously in~\cite{Karananas:2021bqw} in Schr\"{o}dinger CFTs.\,\footnote{Note that Eq.~(3.36) of \cite{Karananas:2021bqw} coincides with \eqref{eq:Dbar_Lambert} after setting $\mu=1$\,, up to slight differences in the conventions used to define various generators. \label{foot:mu_Rsquare}}
This algebraic construction identifies the oscillator Hamiltonian~\eqref{eq:Hosc} as a natural object for characterizing states in the Hilbert space.

Our result adds a natural geometric interpretation to the previous conclusion that we reviewed above. We have shown that the conformal mapping between the TNC geometries leads to the identity~\eqref{eq:state_pic_HamH0} that equates the state-picture Hamiltonian $H_0$ to the operator-picture generators. 
This formula closely resembles the automorphism in Eq.~\eqref{eq:Dbar_Lambert}, but they differ from each other as Eq.~\eqref{eq:state_pic_HamH0} contains the extra contributions from the angular momentum $J$ and the U(1) mass eigenvalue $m$\,.\footnote{This identification also requires setting $\mu=R^2$\,.}
Nevertheless, it is the oscillating part containing $H$ and $C$ that play the key role in the construction of the spectrum, while $J$ and $N$ belong to the stabilizer group of the origin and only contribute a constant shift.
This observation establishes a geometric interpretation of the state-operator correspondence in SU(1\,,\,2) non-Lorentzian CFTs, which was previously formulated more from an algebraic perspective in~\cite{Lambert:2021nol}.

%% file: Sections/Map.tex
\section{A Null-Reduction Derivation of the Conformal Mapping}
\label{sec:state_ope_map}

In Section~\ref{sec:cft_map_su12} we have given a map~\eqref{eq:cmnr} between two TNC geometries with SU(1\,,\,2) conformal Killing vectors in the state and operator pictures, and then used this map to formulate the associated state-operator correspondence.
In this section, we discuss more technical details and provide an explicit derivation of this coordinate transformation from a perspective that is different from the intrinsic approach in Section~\ref{ssec:conf_map_NCgeo}: we will use the four-dimensional ``parent'' backgrounds in the null reduction form~\eqref{eq:nullredmetric} to compute the conformal transformation between the geometries in the state and operator pictures. We will begin in Section~\ref{ssec:requirements_map} by defining the physical constraints on the map. This leads to a set of differential equations which we will solve in Section~\ref{ssec:exact_map}.

\subsection{Constraints on the Conformal Mapping}
\label{ssec:requirements_map}

Our goal is to find a conformal transformation between two metrics with a null isometry. Such metrics take the form of Eq.~\eqref{eq:nullredmetric} and, upon null reduction, are described by the TNC geometries encoded by the background fields given in Eqs.~\eqref{eq:tnc} and \eqref{eq:hthm}, respectively.

We start with the \emph{state picture} and bring the coordinates $(u\,, \, x^0, \, \theta\,, \, \varphi)$ into a more convenient form. Consider the associated TNC geometry~\eqref{eq:tnc}, where we redefine the timelike coordinate $x^0$ to be
\bea
    X^0 = x^0 - \frac{\varphi}{2}\,. 
\eea
The three-dimensional TNC data now become
\begin{subequations} 
\begin{align}
    \tau^{}_\mu \, \dd x^\mu & = R^2  \Bigl[ \dd X^0 + \sin^2 \bigl( \tfrac{\theta}{2} \bigr) \, \dd \varphi \Bigr]\,,
        \qquad%
    h^{}_{\mu\nu} \, \dd x^\mu \, \dd x^\nu = \tfrac{1}{4} \, R^2 \, \bigl( \dd\theta^2 + \sin^2 \theta \, \dd\varphi^2 \bigr)\,, \\[4pt]
    m^{}_\mu \, \dd x^\mu & = \tfrac{1}{2} \Bigl[ \dd X^0 + \cos^2 \bigl(\tfrac{\theta}{2}\bigr) \, \dd \varphi \Bigr]\,,
\end{align}
\label{eq:NCdata_state_sec4}
\end{subequations}
while the four-dimensional metric~\eqref{eq:nullredmetric} incorporating the null coordinate $u$ becomes
\beq
\label{eq:nullredmetric_state4}
    \dd s^2 
    = R^2 \, \biggl\{ 2 \, \dd u \, \Bigl[ \dd X^0 + \sin^2 \bigl(\tfrac{\theta}{2}\bigl) \, \dd \varphi \Bigr] - \bigl(\dd X^0\bigr)^2 - \dd X^0 \, \dd \varphi + \tfrac{1}{4} \, \dd \theta^2 \biggr\}\,.
\eeq
Next, we consider the \emph{operator picture} with geometry~\eqref{eq:hthm}, and define
\bea
    \mathbf{u} = \hat{u} + \frac{\hat{\varphi}}{2}\,. 
\eea
The three-dimensional TNC data in~\eqref{eq:hthm} are mapped to
\beq
    \hat{\tau}^{}_\mu \, \dd x^\mu  = \dd \hat{t} + \hat{r}^2 \, \dd \hat{\varphi} \,,
        \qquad%
    \hat{h}^{}_{\mu\nu} \, \dd x^\mu \, \dd x^\nu  = \dd \hat{r}^2 + \hat{r}^2 \, \dd \hat{\varphi}^2 \,, 
        \qquad%
    \hat{m}^{}_\mu \, \dd x^\mu  = \tfrac{1}{2} \, \dd \hat{\varphi}   \,.
\label{eq:NCdata_ope_sec4}
\eeq
The corresponding four-dimensional metric in the null-reduction form is
\begin{equation}
\label{eq:nullredmetric_ope4}
    \dd \hat{s}^2  
    = 2 \, \dd \mathbf{u} \, \bigl( \dd \hat{t} + \hat{r}^2 \, \dd \hat{\varphi} \bigr) + \dd \hat{r}^2 - \dd \hat{t} \, \dd \hat{\varphi}\,.
\end{equation}
Our task now is to determine the conformal transformation between the line elements~\eqref{eq:nullredmetric_state4} and \eqref{eq:nullredmetric_ope4}.
It is natural to require that the following criteria are satisfied:
\begin{itemize}
    \item The null directions $u$ in the state picture and $\mathbf{u}$ in the operator picture should correspond to the same null isometry, \emph{i.e.}, 
    \beq
    \p_u = \p_{\mathbf{u}}\, ,
    \label{eq:preserving_null_isometry}
    \eeq
    which implies the null momenta $P_u$ and $P_{\mathbf{u}}$ in two pictures are identified with each other and should not be affected by the map.\,\footnote{In this regard, notice that the shift $\hat{u} = \mathbf{u} - \hat{\varphi}/2$ does not affect the null momentum.} 
    This implies that the coordinate transformation between the two geometries should take the following form:
    \beq
    u = \mathbf{u} + \mathbf{F}\bigl(\hat{t}\,,\,\hat{r}\,,\,\hat{\varphi}\bigr) \, ,  
    \label{eq:transf_u_uhat}
    \eeq
    where $\mathbf{F}$ is analytic and transformations of the other coordinates are independent of 
    $\mathbf{u}$\,. 
    
    \item We require that the compact angular coordinates $\varphi$ and $\hat{\varphi}$ correspond to the same isometry, \emph{i.e.},
    \beq
    \p_{\varphi} = \p_{\hat{\varphi}}  \, .
    \label{eq:preserving_ang_isometry}
    \eeq
    This implies that the coordinate transformation between the two coordinate systems should be of the form
    \beq
        \varphi = \hat{\varphi} + \mathbf{G}\bigl(\hat{t}\,,\,\hat{r}\bigr) \, , 
    \eeq
where $\mathbf{G}$ is analytic and transformations of the other coordinates are independent of $\hat{\varphi}$, including the function 
$\mathbf{F}$ introduced in Eq.~\eqref{eq:transf_u_uhat}.

\end{itemize}
Combining these two requirements leads to the following ansatz for a coordinate transformation between the metrics in the state and operator pictures:
\beq
 u = \mathbf{u} + \mathbf{F}\bigl(\hat{t}\,,\,\hat{r}\bigr) \, , \qquad
    X^0 = X^0\bigl(\hat{t}\,,\,\hat{r}\bigr)\,,  \qquad
  \theta =  \theta\bigl(\hat{t}\,,\,\hat{r}\bigr)\,, \qquad
    \varphi = \hat{\varphi} + \mathbf{G}\bigl(\hat{t}\,,\,\hat{r}\bigr) \,.
\label{eq:ansatz_map_sec4}
\eeq
Moreover, we allow for any Weyl rescaling that preserves the form of the null-reduction metric~\eqref{eq:nullredmetric} and that is independent of the coordinates $u$ and $\varphi$ (or, analogously, $\mathbf{u}$ and $\hat{\varphi}$).
In other words, we require
\beq
    \dd s^2 = \mathbf{\Omega}^2 \bigl(\hat{t}\,,\, \hat{r}\bigr) \, \dd\hat{s}^2\,.        
\label{eq:matching_conformal}
\eeq 
Note that the above four-dimensional Lorentzian formalism with a null isometry is equivalent to the three-dimensional formalism, where one asks about how the TNC data $(\tau\,, \,m\,,\, h_{\mu\nu})$ and $(\hat{\tau}\,,\, \hat{m}\,, \, \hat{h}_{\mu\nu})$ are mapped to each other, up to Weyl rescaling, local Galilean boosts and $\mathrm{U}(1)$ gauge transformations.

\subsection{Solving the Constraints}
\label{ssec:exact_map}

We now compute the exact mapping between the geometries \eqref{eq:nullredmetric_state4} and \eqref{eq:nullredmetric_ope4}, which we write compactly as, respectively, 
\begin{align}
    \text{\emph{state picture:}\quad}
    \dd s^2 = g^{}_\text{MN} \, \dd x^\text{M} \, \dd x^\text{N}\,;
        \qquad%
    \text{\emph{operator picture:}\quad}
    \dd \hat{s}^2 = \hat{g}^{}_\text{MN} \, \dd \hat{x}^\text{M} \, \dd \hat{x}^\text{N}\,.
\end{align}
Here, ``M'' denotes the four-dimensional index. Then, the Weyl condition~\eqref{eq:matching_conformal} implies
\bea \label{eq:wcst}
    g^{}_\text{MN} \, \frac{\p x^\text{M}}{\p \hat{x}^\text{K}} \, \frac{\p x^\text{N}}{\p \hat{x}^\text{L}} = \mathbf{\Omega}^2 \bigl( \hat{t}\,, \, \hat{r} \bigr) \, \hat{g}^{}_\text{KL}\,. 
\eea
Plugging the ansatz~\eqref{eq:ansatz_map_sec4} into the Weyl condition~\eqref{eq:wcst} gives rise to a set of differential equations. In terms of $\mathbf{t} \equiv \hat{t} / \bigl(2 \, R^2\bigr)$ and $\mathbf{r} \equiv \hat{r} / R$ as defined in Eq.~\eqref{eq:bold_tr}, together with $\mathbf{\Theta} \equiv \theta / 2 \in [0\,, \, \pi/2]$\,, we write these differential equations as
\begin{subequations} \label{eq:equation-all}
\begin{align} \label{eq:all-1}
    \mathbf{\Omega} &= \frac{\sin \mathbf{\Theta}}{2 \, \mathbf{r}}\,, 
        &%
    \p_\mathbf{t} X^0 &= 2 \, \sin^2 \mathbf{\Theta} \, D_\mathbf{t} \mathbf{F}\,, 
        &%
    \bigl( \p_\mathbf{t} \mathbf{\Theta} \bigr)^2 + \sin^2\bigl(2\mathbf{\Theta}\bigr) \, \bigl( D_\mathbf{t} \mathbf{F} \bigr)^{2} = \frac{\sin^2 \mathbf{\Theta}}{4 \, \mathbf{r}^4}\,,& \\[4pt] \label{eq:all-2}
    \mathbf{G} &= - 2 \, \mathbf{F}\,,
        &%
    \p_\mathbf{r} X^0 &= 2 \, \sin^2 \mathbf{\Theta} \, \p_\mathbf{r} \mathbf{F}\,, 
        &%
    \bigl( \p_\mathbf{r} \mathbf{\Theta} \bigr)^2 + \,
    \sin^2\bigl(2\mathbf{\Theta}\bigr) \,\, \bigl( \p_\mathbf{r} \mathbf{F} \bigr)^2 = \frac{\sin^2 \mathbf{\Theta}}{\mathbf{r}^2}\,,& \\[8pt] \label{eq:all-3}
        &&&& 
    \p_\mathbf{t} \mathbf{\Theta} \, \p_\mathbf{r} \mathbf{\Theta} + \sin^2\bigl(2\mathbf{\Theta}\bigr) \, D_\mathbf{t} \mathbf{F} \, \p_\mathbf{r} \mathbf{F} = 0\,,&
\end{align}
\end{subequations}
where we have defined
$D_\mathbf{t} \mathbf{F} \equiv \p_\mathbf{t} \mathbf{F} + (4 \, \mathbf{r}^2)^{-1}$\,.

In order to solve for $\mathbf{\Theta}$ and $\mathbf{F}$\,, it is natural to take the following reparametrizations of their derivatives, that are compatible with Eq.~\eqref{eq:equation-all}, in terms of a new angle $\beta$\,:
\begin{subequations} \label{eq:fdtf}
\begin{align}
	\partial_\mathbf{r} \mathbf{\Theta} &= \frac{1}{\mathbf{r}} \, \sin \mathbf{\Theta} \, \cos\beta\,, 
        &%
    \sin \bigl(2 \mathbf{\Theta} \bigr) \, \partial_\mathbf{r} \mathbf{F} &= -\frac{1}{\mathbf{r}} \, \sin \mathbf{\Theta} \, \sin\beta\,, \\[4pt]
    \partial_\mathbf{t} \mathbf{\Theta} &= \frac{1}{2 \, \mathbf{r}^2} \, \sin \mathbf{\Theta} \, \sin\beta\,, 
        &%
    \sin \bigl(2 \mathbf{\Theta}\bigr) \,   D_{\mathbf{t}} \mathbf{F} &= \frac{1}{2 \, \mathbf{r}^2} \, \sin \mathbf{\Theta} \, \cos\beta\,,
\end{align}
\end{subequations}
Eliminating $\beta$ in Eq.~\eqref{eq:fdtf}, we find
\begin{align}\label{eq:reducedequation}
\begin{split}
    \partial_\mathbf{r} \mathbf{\Theta}  = 2 \, \mathbf{r} \, \sin \bigl(2 \mathbf{\Theta}\bigr) \, D_{\mathbf{t}} \mathbf{F} , 
        \qquad%
    \partial_\mathbf{t} \mathbf{\Theta} = - \frac{\sin (2 \mathbf{\Theta})}{2 \, \mathbf{r}} \, \p_\mathbf{r} \mathbf{F}\,.
\end{split}
\end{align}
Finally, in terms of $\mathbf{W} =\ln \tan  \mathbf{\Theta}$\,, we rewrite Eq.~\eqref{eq:reducedequation} as the following decoupled second-order partial differential equations:
\begin{align}
    \frac{\p^2 \mathbf{F}}{\p \mathbf{t}^2} + 
\frac{1}{4 \, \mathbf{r}} \frac{\partial}{ \p \mathbf{r}} \left( 
\frac{1}{\mathbf{r}} \frac{\p \mathbf{F}}{\p \mathbf{r}}
\right) &= 0\,, 
        &%
    \frac{\p^2 \mathbf{W}}{\p \mathbf{t}^2} +\frac{1}{4 \, \mathbf{r}} \frac{\partial}{ \p \mathbf{r}} \left( 
\frac{1}{\mathbf{r}} \frac{\p \mathbf{W}}{\p \mathbf{r}} -\frac{1}{\mathbf{r}^2}
\right) &=0\,,
\end{align}
to which we find the general solutions, 
\begin{equation}\label{eq:thetafasD1D2}
\mathbf{\Theta}(\mathbf{t},\mathbf{r}) = \arctan \! \left[ 
\frac{c\, \mathbf{r}}{\sqrt{D_1(U_+) \, D_2(U_-)}} \right], 
    \qquad%
\mathbf{F} (\mathbf{t} ,\mathbf{r} ) = -\frac{i}{4} \ln \! \left[\frac{ D_1(U_+)}{ D_2(U_-)} \right] .
\end{equation}
Here, $c$ is an integration constant and we have introduced two arbitrary functions $D_1(U)$ and $D_2(V)$ with $U_\pm = \mathbf{t} \pm i\, \mathbf{r}^2$\,, which are subject to the constraints 
\begin{equation}\label{eq:reorga-constraint-0}
    \frac{- i \, c^2}{D_1'(U_+) \, D_2'(U_-)} + \frac{D_2(U_-)}{D_2'(U_-)} - U_- 
        = 
    \frac{D_1(U_+)}{D_1'(U_+)} - U_+\, .
\end{equation}
Differentiating Eq.~\eqref{eq:reorga-constraint-0} with respect to $U_+$ and $U_-$\,, respectively, we find 
\bea
    D''_1(U_+) = D''_2(U_-) = 0\,,
\eea
\emph{i.e.}~the only solution to the constraint~\eqref{eq:reorga-constraint-0} is that both $D_1(U_+)$ and $D_2(U_-)$ are linear functions. 

Therefore, up to integration constants, which qualitatively lead to the same form for the change of coordinates, we find the following unique real solutions to the above differential equations: 
\begin{subequations} \label{eq:formap}
\begin{align} 
    X^0 & =  - \mathrm{arccot} \!  \le \frac{\mathbf{t}^2 + \mathbf{r}^4 - 1}{2 \, \mathbf{t}} \ri,  
        &%
    \mathbf{\Theta} &= \arctan \frac{2 \, \mathbf{r}}{\sqrt{\mathbf{t}^2 + (\mathbf{r}^2-1 )^{2}}}\,, \\[4pt]
    \varphi &= \hat{\varphi} - \mathrm{arctan}\! \le \frac{\mathbf{r}^2 - 1}{\mathbf{t}} \ri, &   
u &= \mathbf{u} + \frac{1}{2} \, \mathrm{arctan} \! \le \frac{\mathbf{r}^2 - 1}{\mathbf{t}} \ri . %
\label{eq:map}
\end{align}
\end{subequations}
The above form is associated with the choices $D_1(U_+) = U_+-i$ and $D_2(U_-) = U_-+i$\,. Note that a more general set of solutions can be trivially generated by performing the $z=2$ Lifshitz rescaling,
\bea
    \mathbf{t} \rightarrow b^2\,\mathbf{t}\,,
        \qquad%
    \mathbf{r} \rightarrow b\,\mathbf{r}\,.
\eea
Plugging $X^0 = x^0 - \varphi/2$\,, $\mathbf{\Theta} = \theta/2$ and $\mathbf{u} = \hat{u} + \hat{\varphi}/2$ into the conformal mapping~\eqref{eq:formap}, we finally derive Eq.~\eqref{eq:cmnr}. We also find that the inverse map is given by
\begin{subequations}
\begin{align}
    \mathbf{t} & =  \frac{2 \, \sin X^0 \, \cos \mathbf{\Theta}}{\bigl(\sin X^0 \bigr)^{\!2} + \bigl( \cos X^0 + \cos \mathbf{\Theta} \bigr)^{\!2}}\,,
        &%
    \hat{\varphi} &= \varphi -  \mathrm{arctan} \! \le \frac{\cos X^0 + \cos \mathbf{\Theta} }{\sin X^0}   \ri,   \\[4pt]
    \mathbf{r} & =  \frac{\sin \mathbf{\Theta}}{\sqrt{\bigl(\sin X^0 \bigr)^{\!2} + \bigl( \cos X^0 + \cos \mathbf{\Theta} \bigr)^{\!2}}}\,,
        &%
    \mathbf{u} &= u  + \frac{1}{2} \,   \mathrm{arctan}  \! \le \frac{\cos X^0 + \cos \mathbf{\Theta} }{\sin X^0} \ri.
\label{eq:inverse_map}
\end{align}
\end{subequations}

Note that, in the $R \rightarrow \infty$ limit, the transverse sphere becomes non-compact, \ie $S^2 \rightarrow \mathbb{R}^2$\,. In this limit, the four-dimensional manifold is the null reduction of the Lorentzian plane. Therefore, the state and operator metrics are now related to each other by a diffeomorphism, and the Weyl factor reduces to $\mathbf{\Omega} = 1$\,. Moreover, at the leading order in $R^{-1}$ the TNC data in Eqs.~\eqref{eq:NCdata_state_sec4} and \eqref{eq:NCdata_ope_sec4} coincide without any extra local Galilean boost \emph{nor} gauge transformation.\,\footnote{This statement does \emph{not} hold at higher orders in $R^{-1}$\,. In general, additional local Galilean boosts of the TNC data are required in order to match the line elements in the state and operator pictures.}
Namely, under the expansion at leading-order in $\mathbf{t}$ and $\mathbf{r}$ of the map~\eqref{eq:map},\footnote{Due to Eq.~\eqref{eq:bold_tr}, an expansion for large radius $R$ of the three-sphere corresponds to small $\mathbf{t}$ and $\mathbf{r}$\,.}
\begin{subequations}
\begin{align}
    X^0 &= 2 \, \mathbf{t} + \mathcal{O}\bigl(\mathbf{t}^3\bigr)\,,
        &%
    \varphi &= \hat{\varphi} + \frac{\pi}{2} + \mathcal{O}\bigl(\mathbf{t}\bigr)\,, \\[4pt]
    u &= \mathbf{u} - \frac{\pi}{4} + \mathcal{O}\bigl(\mathbf{t}\bigr)\,,
        &%
    \theta &= 4 \, \mathbf{r}  + \mathcal{O}\bigl(\mathbf{r}^3\bigr)\,,
\end{align}
\label{eq:map_leading_order}
\end{subequations}
we are led to
\beq
h = \hat{h}  + \mathcal{O}\bigl(R^{-1}\bigr)  \, , \qquad
\tau = \hat{\tau}  + \mathcal{O}\bigl(R^{-1}\bigr)  \, , \qquad 
m = \hat{m}  + \mathcal{O}\bigl(R^{-1}\bigr) 
\eeq
for the TNC data in Eqs.~\eqref{eq:tnc} and \eqref{eq:hthm}.

%% file: Sections/Discussion.tex
\section{Discussion}
\label{sec:discussions}

In this paper, we formulated a state-operator correspondence for three-dimensional non-Lorentzian CFTs with $\mathrm{SU}(1,2) \times \mathrm{U}(1)$ symmetry group.

Firstly, we derived a state-picture background by performing a dimensional reduction along a null direction $u$ of the Lorentzian cylinder $\mathbb{R} \times S^3$.
The null reduction is chosen to satisfy two criteria:
the energy $E= i \partial_{x^0}$ is associated with a globally defined time coordinate $x^0$, and the null momentum $P_u$ is selected such that 
the centralizer of $P_u$ in the set of conformal Killing vectors in the parent geometry gives rise to the SU(1,\,\,2) symmetry group.
Secondly, we found the operator-picture background from an infinite radius limit of the state-picture metric.
This background belongs to a class of manifolds with SU(1,\,\,2) conformal symmetry, obtained in \cite{Lambert:2021nol} from an $\Omega$-deformed null reduction of Minkowski spacetime.
Finally, we determined an exact conformal map relating the two geometries. This conformal map preserves the null isometry, which is essential to our construction.

Under the conformal map, the Hamiltonian generating the evolution between equal-time slices in the state picture is mapped to a linear combination of generators in the operator picture (see Table~\ref{tab:intro_results}).
This mapping provides a geometric interpretation for the results obtained previously via algebraic approaches~\cite{Nishida:2007pj,Lambert:2021nol,Karananas:2021bqw}.

\begin{table}[b!]   
\begin{center}    
\begin{tabular}  {c|c|c}   & \textbf{state picture}  & \textbf{operator picture}  \\ \hline
\rule{0pt}{4.9ex}\textbf{Lorentzian CFT}    &  $H^\text{L}_0 = \, \p^{}_\tau$ & $\frac{1}{2} \le P_0 + K_0 \ri$  \\
\rule{0pt}{4.9ex} \textbf{SU(1$\,\textbf{,}\,$2) CFT}     &  $H_0 = \p^{}_{x^0}$ & $\frac{1}{2} \Bigl( R^2 \, H + \frac{C}{R^2} - J - N \Bigr) $  
\end{tabular}  
\caption{Map between the Hamiltonian $H_0$ in the state picture (left column) and generators in the operator picture. Here, $P_0$ and $K_0$ denote the temporal component of the momentum and of the special conformal generators, respectively. Moreover, $H$ is the Hamiltonian, $C$ the generator of special conformal transformations, $J$ the angular momentum, $N$ the particle number generator, and $R$ the radius of the three-sphere.} 
\label{tab:intro_results}
\end{center}
\end{table}

We discuss implications of our results for field theories and holography below. 

\begin{center}
--- \textbf{Field-Theoretical Aspects} ---
\end{center}

\noindent $\bullet$~\emph{Dilatation operator in the SU(1\,,\,2) state-operator correspondence.}
The geometric map derived in this paper establishes the relation between the state-picture Hamiltonian $H_0$ and generators in the operator picture (see Table~\ref{tab:intro_results}).
In the Lorentzian case, it is well known that the Euclidean dilatation operator $D^{\rm E}$ can be mapped to the Euclidean Hamiltonian $H^{\rm E}$ in the state-picture via a similarity transformation, modulo an analytic continuation (see Eq.~\eqref{eq:rel_automorphism}). 
It would be interesting to understand whether a similar interpretation exists for SU(1\,,\,2) CFTs. 
In imaginary time, we define a dilatation operator $\bar{D}$ in the operator picture and a Hamiltonian $\bar{H}$ in the state picture, such that
\beq
\bar{D} \equiv -  \frac{i}{2} \le  R^2 \, H + \frac{C}{R^2} - J -m \ri, \qquad
\bar{H} \equiv  -i H_{0} \, ,
\label{eq:Euclidean_dilatation_su12}
\eeq
These operators are in an anti-Hermitian representation. 
A natural expectation would be that the non-Lorentzian SU(1\,,\,2) state-operator correspondence should equate $\bar{D}$ to $\bar{H}$.
This correspondence is hinted by the observation that the eigenvalues of the state-picture Hamiltonian $H_0$ correspond to the scaling dimensions of the local operators, similarly as in Eq.~\eqref{eq:Hosc}. 
However, the precise coordinate transformation that maps the generators to $\bar{D}$ is obscured by how a Wick-like rotation could be defined in non-Lorentzian quantum field theories, which in general involve Galilean boosts and thus a non-standard analytic continuation. We leave this investigation for future studies.
    \vspace{3mm} 
    
\noindent $\bullet$~\emph{Generalized state-operator correspondences.}
The techniques used in this paper for building the state-operator correspondence between geometries with SU$(1,2)$ symmetry may be generalized to CFTs on other manifolds. Usually, it is difficult to build a state-operator correspondence for a CFT on a generic background. For instance, it is argued in~\cite{Belin:2018jtf} that such a correspondence might not even exist on the torus. It is therefore interesting to ask what symmetry principles are required for the existence of a correspondence similar to the one discussed in our paper. One possibility is that CFTs with generalized symmetries (like higher-form symmetries) might lead to a state-operator correspondence involving non-local operators \cite{Hofman:2024oze}. Another promising direction is to investigate the existence of a state-operator correspondence for $l$-conformal Galilei algebra \cite{Bagchi:2009my,henkel1997local,negro1997nonrelativistic,Masterov:2023owu} and the conformal Carroll algebra \cite{Duval:2014uva,Bagchi:2019xfx}. In these symmetry group, the temporal special conformal transformations survive as in the SU($1,2$) case that we have discussed.

\vspace{3mm}

\noindent $\bullet$~\emph{Null reduction of $\mathcal{N}=4$ super Yang-Mills.}
    The geometries determined in this work provide a natural background to couple non-Lorentzian CFTs invariant under the $\mathrm{SU}(1,2) \times \mathrm{U}(1)$ group.  
    The null reduction of free theories with such symmetry was computed on the $\Omega$-deformed spacetime in~\cite{Lambert:2021nol,Smith:2023jjb}.
    It would be interesting to extend this analysis by null reducing the $\mathcal{N}=4$ SYM action, including the interactions. 
    It would be important to identify a null momentum generator $P_u$ inside the $\mathrm{PSU}(2,2|4)$ group, such that the set of generators commuting with $P_u$ form the $\mathrm{PSU}(1,2|3)$ subgroup.
    This subgroup plays an important role as it underlies interesting dual black hole solutions~\cite{Gutowski:2004yv}. It would be interesting to understand the microscopic nature of such black hole solutions using the $\mathrm{PSU}(1,2|3)$  dual field theory.\,\footnote{Complementarily, for a non-exhaustive list of recent developments on the microstates of these black holes using indices or other techniques, see~\cite{Grant:2008sk,Choi:2018hmj,Benini:2018ywd,Murthy:2020scj,Goldstein:2020yvj,Aharony:2021zkr,Chang:2022mjp,Cabo-Bizet:2018ehj}.}

    \vspace{3mm}

\noindent $\bullet$~\emph{Relation to Spin Matrix Theories.}
    A concrete example of SU$(1,\,2)$-invariant models comes from the Spin Matrix Theories (SMTs)~\cite{Harmark:2014mpa,Baiguera:2023fus}.
    The interacting Hamiltonians of SMTs have been determined as a quantum mechanical theory~\cite{Harmark:2019zkn,Baiguera:2020jgy,Baiguera:2020mgk,Baiguera:2021hky,Baiguera:2022pll}.
    In particular, the SU$(1,2)$-invariant SMTs can be viewed as (2+1)-dimensional quantum field theories. A more interesting SMT is the one with the largest symmetry group $\mathrm{PSU}(1,2|3)$. 
    It would be interesting to realize such a non-Lorentzian superymmetric CFT coupled to the geometries discussed in this work.

 \vspace{3mm} 

\noindent $\bullet$~\emph{SU(1,2) as a Killing symmetry}. 
    The SU$(1,2)$ symmetry can be realized as the Killing symmetry of a quaternionic K\"ahler manifold in four dimensions. 
    There are different constructions of coordinate systems that describe the SU$(1,2)$ Killing symmetry: complex $c$-map, Iwasawa, and Harish-Chandra coordinates  \cite{Gunaydin:2007qq}.
    These coordinate systems make manifest different structures of the quaternionic K\"ahler manifold.
    In particular, the differential representations of the Killing vectors in Iwasawa and complex $c$-map coordinates are very closely related to the conformal Killing vectors of a null-reduced background. 
    On the contrary, the differential representation in Harish-Chandra coordinates fits better with the algebraic structure of SMT. 
    These analogies may shed light on the relations between different bases in SMT and the null reduced geometry.

\begin{center}
    --- \textbf{Holographic Aspects} ---
\end{center}

\noindent $\bullet$~\emph{Top-down perspective from string theory.}
In the context of holography, the results of this paper focus on the field-theoretical side and are related to null-reducing $\mathcal{N} = 4$ SYM. This line of thinking naturally fits within the program of mapping out BPS decoupling limits of string theory recently developed in~\cite{Blair:2023noj, Gomis:2023eav,Blair:2024aqz, Lambert:2024ncn}.\,\footnote{See also a complementary perspective using BPS mass spectra in~\cite{bpslimits}.} 
An embedding within string theory might be central for us to make sense out of the validity of the null reduction of $\mathcal{N} = 4$ SYM. 
In a T-dual frame, $\mathcal{N} = 4$ SYM on a null compactification gives rise to a field theory on D2-branes in a corner of type IIA superstring theory. Such a corner supposedly arises from taking two consecutive BPS decoupling limits of the IIA theory, such that we zoom in on a background bound-state of a fundamental string and a D2-brane~\cite{Gomis:2023eav,Harmark:2025ikv}. In flat spacetime, this double BPS decoupling limit leads to a Galilean version of SYM. However, in order for the SU($1,2$) symmetry to arise, such that the theory is conformal, it is necessary to consider the curved TNC geometries with an $\Omega$-deformation, as we have discussed in the current paper. It would be interesting to understand the twists introduced by the $\Omega$-deformation on both the field theory side and the dual bulk gravity, which is associated with a null reduction of the AdS${}_5$ geometry. Furthermore, it would be interesting to take one step further and study how SMTs arise from a further BPS decoupling limit. 

\vspace{3mm} 

\noindent $\bullet$~\emph{Relation to black hole microstates.}
The discussions that we have had on the field theory side may eventually help us understand the black hole microscopic states. First, in the state picture, via holography, it is expected that solutions to the ground state equation of the $\mathrm{PSU}(1,2|3)$ SMT Hamiltonian capture the microscopic states of the $1/16$-BPS black hole in AdS$_5$~\cite{Baiguera:2022pll}.  
Second, in the operator picture, recent works~\cite{Chang:2022mjp,Choi:2022caq,Choi:2023znd,Choi:2023vdm, Chang:2024zqi} showed that the operators in $\mathcal{N}=4$ SYM can be classified into the so-called fortuitous and monotone operators. It is then conjectured in~\cite{Chang:2024zqi} that, via the AdS/CFT correspondence, the fortuitous operators are dual to typical black hole microstates, while the monotone ones are dual to smooth horizonless geometries. It would be interesting to use the state-operator correspondence developed in the current paper to link these two algorithms.

Another potentially interesting way to understand the microscopic states of AdS black holes using non-Lorentzian CFTs is related to the works~\cite{Dorey:2022cfn, Dorey:2023jfw, Mouland:2023gcp},
where a precise matching is made between the entropy of AdS$_7\times S^4$ black holes in the Penrose limit and the superconformal quantum mechanics on the moduli space of Yang-Mills instantons. 
This non-relativistic quantum mechanical system arises from the discrete light cone quantization (DLCQ) of the six-dimensional (2,0) theory~\cite{Maldacena:2008wh}. In connection to the current paper, it would be interesting to generalize these methods to SU($1,n$) field theories.

%% file: Sections/App_su12algebra.tex
\section{Bases of the \texorpdfstring{$\mathfrak{su}(1,2)$}{su(1,2)} Algebra}
\label{app:su12_algebra}

There are various bases that are used in the literature to represent the $\mathfrak{su}(1,2)$ algebra~\cite{Gunaydin:2007qq,Biedenharn, bars1990unitary} (see also Appendix~C of~\cite{Baiguera:2020mgk} and Section~2 of \cite{Baiguera:2022pll} for further details). 
The first one, reported in Eq.~\eqref{eq:suot}, naturally arises 
from the conformal Killing vectors of the $\Omega$-deformed background that we have discussed in Section~\ref{ssec:su12_algebra}.
The second one, that we discuss below, makes the underlying Schr\"{o}dinger group structure manifest. 
The two bases are related via the automorphism~\eqref{eq:automorphismsu12}. 

A representation of the $\mathfrak{su}(1,2) $ algebra relevant to the operator picture naturally arises in Schr\"{o}dinger CFTs. 
To find it, we start with the $d$--dimensional relativistic conformal algebra $\mathfrak{so}(2,4)$ in Minkowski space, and then perform a null reduction along the light-cone direction $\hat{x}^+$ followed by an $\Omega$-deformation~\cite{Lambert:2021nol} (see also Appendix~\ref{app:ssec:coord_ope} for further details).
The $\mathfrak{su}(1,2)$ algebra is then obtained as the subset of generators of $\mathfrak{so}(2,4)$ that commute with the null momentum $\mathcal{P}_+$.
To this aim, let us recall the conformal algebra,  
\begin{subequations} \label{eq:ifactor}
\begin{align}
    \bigl[\mathcal{D}, \mathcal{P}^{}_\text{M}\bigr] &=  -  \mathcal{P}^{}_\text{M} \, , 
        &%
    \bigl[\mathcal{J}^{}_\text{MN}, \mathcal{P}^{}_\text{L}\bigr] &=   \eta^{}_\text{ML} \, \mathcal{P}^{}_\text{N} - \eta^{}_\text{NL} \, \mathcal{P}^{}_\text{M}  \, ,  \\[4pt]
    \bigl[\mathcal{D}, \mathcal{K}^{}_\text{M}\bigr] &=  \mathcal{K}_\text{M} \, , 
        &%
    \bigl[\mathcal{J}^{}_\text{MN}, \mathcal{K}_\text{L}\bigr] &= \eta^{}_\text{ML} \, \mathcal{K}^{}_\text{N} - \eta^{}_\text{NL} \, \mathcal{K}^{}_\text{M} \, ,  \\[4pt]
   \bigl[\mathcal{K}^{}_\text{M}, \mathcal{P}^{}_\text{N}] &= 2 \, \bigl(\eta^{}_\text{MN} \, \mathcal{D} + \mathcal{J}^{}_\text{MN} \bigr)\, , 
        &%
    \bigl[\mathcal{J}^{}_\text{MN}, \mathcal{J}^{}_\text{KL}] & = \eta^{}_\text{NL} \, \mathcal{J}^{}_\text{MK} -  \eta^{}_\text{NK} \, \mathcal{J}^{}_\text{ML} + (\text{M} \leftrightarrow \text{N})  \, ,
\end{align}
\end{subequations}
where $\eta_\text{MN} = \mathrm{diag}(-1,1, 1, 1)$.
The set of generators is composed by the momentum $\mathcal{P}_\text{M}$\,, the 
rotations $\mathcal{J}_\text{MN}$\,, the dilatation operator $\mathcal{D}$ and the special conformal generators $\mathcal{K}_\text{M}$\,.
Their differential representation is given by
\begin{subequations} \label{eq:diff_representation_conf_gen}
\begin{align}
    \mathcal{P}_\text{M} &=  - \hat{\p}_\text{M} \, , 
        &%
    \mathcal{J}_{MN} &= -  \hat{x}_\text{M} \, \hat{\p}_\text{N} + \hat{x}_\text{N} \, \hat{\p}_\text{M}   \, , \\[4pt]  
    \mathcal{D} &= \hat{x}^\text{M} \, \hat{\p}_\text{M} \, , 
        &%
    \mathcal{K}_\text{M} &=  - \hat{x}_\text{N} \, \hat{x}^\text{N} \, \hat{\p}_\text{M} + 2 \hat{x}_\text{M} \, \hat{x}^\text{N} \, \hat{\p}_\text{N}  \, . 
\end{align}
\end{subequations}
The subalgebra of the conformal generators of Minkowski space that commute with the null momentum $\hat{\partial}_+$ can be represented by differential operators.
In light-cone coordinates of Minkowski spacetime, the generators spanning the associated $\mathfrak{su}(1,2) \oplus \mathfrak{u}(1)$ algebra are given by \cite{Lambert:2021nol}  
\begin{subequations}
\label{eq:differ-lambert}
\begin{align}
    N &= 2 \, \partial_+, 
        \quad%
    H = - \partial_-, 
        \quad%
    P_i = - \partial_i - \frac{\epsilon_{ij} \, x^j}{2} \partial_-, 
        \quad%
    J= \epsilon^{ij} \, x_j \, \p_i,
        \quad%
    D= 2\,x^-\partial_- \!+ x^i \partial_i, \\[4pt]
    G_i &= x^i \biggl( \p_u - \frac{x_k \, x^k}{8} \p_t - \frac{\epsilon_{j}{}^k \, x^j}{2} \p_k \biggr) + x^- \p_i + \epsilon_{ij} \biggl[ \frac{x^j}{2} \, \bigl( x^k \, \p_k + x^- \, x^j \p_t \bigr) - \frac{x_k \, x^k}{4} \p_j \biggr], \\[4pt]
    C &= -\frac{x_j \, x^j}{2} \, \partial_+ - \left[ 
(x^-)^2- \biggl(\frac{x_j \, x^j}{4} \biggr)^{\!\!2}\,
\right] \partial_- - \biggl(
\frac{x_k \, x^k}{4} \, \epsilon_{ij} \, x_j +x^- x^i
\biggr) \, \partial_i\,,
\end{align}
\end{subequations}
where we specialized to an $\Omega$-deformation of the form $\Omega_{ij} =  \epsilon_{ij}$, since we are working in two spatial dimensions (see Appendix~\ref{app:ssec:coord_ope} for more details on the null reduction).
These are the conformal Killing vectors of Minkowski spacetime which commute with the translation $\mathcal{P}_+$ along the null direction 
\cite{Lambert:2021nol}.

Changing coordinates to the hatted ones in Eq.~\eqref{eq:hthm}, the generators become 
\begin{subequations}\label{eq:ourcoordinateSU12}
\begin{align}
    H &= 2\, \partial_{\hat{t}}\,, 
        \qquad%
    J= -\partial_{\hat{\varphi}}\,, 
        \qquad%
    D= 2 \, \hat{t} \, \partial_{\hat{t}} + \hat{r} \, \partial_{\hat{r}}\,,
        \qquad%
    N = \partial_{\hat{u}}\,, \\[4pt]
    C &=  -\frac{\hat{r}^2}{4} \partial_{\hat{u}} + \frac{1}{2}  \left(\hat{t}^{\,2} -\frac{\hat{r}^4}{4} \right) \partial_{ \hat{t}} + \frac{\hat{t} \, \hat{r} }{2} \partial_{\hat{r}}
+\frac{\hat{r}^2}{4} \partial_{ \hat{\varphi}} \, , \\[4pt]
    P &\equiv \frac{P_1+i P_2}{\sqrt{2}} = - \frac{e^{i\hat{\varphi}}}{\sqrt{2}} \, \Bigl( i \, \hat{r} \,   \partial_{\hat{t}} + \partial_{\hat{r}} + i \, \hat{r}{}^{-1} \partial_{\hat{\varphi}} \Bigl)\,, \\[4pt]
    G& \equiv \frac{G_1+i G_2}{\sqrt{2}} =\frac{e^{i\hat{\varphi}}}{4\sqrt{2}} \biggl[ 2 \, \hat{r} \, \partial_{\hat{u}} + \bigl( 
\hat{r}^2 - 2 \, i \, \hat{t}
\, \bigr) \bigl( \hat{r} \, \partial_{\hat{t}} - i \, \partial_{\hat{r}} \bigr) - \bigl( 3 \, \hat{r}^2 + 2 \, i \, \hat{t} \, \bigr) \frac{\partial_{\hat{\varphi}}}{\hat{r}}
    \biggr]\,.
\end{align}
\end{subequations}
Both the generators in \eqref{eq:differ-lambert} and  \eqref{eq:ourcoordinateSU12} span the $\mathfrak{su}(1,2) \oplus \mathfrak{u}(1)$ algebra \eqref{eq:msuot} and \eqref{eq:sa}. 
The $\mathfrak{su}(1,2) \oplus \mathfrak{u}(1)$ algebra in the new basis \eqref{eq:msuot} and \eqref{eq:sa} is related to the Schr\"odinger algebra by a contraction. Indeed, let us rescale the generators $P_i,G_i$ by a factor $\epsilon$ as \eqref{eq:rescaling-PG}.
The Schr\"odinger algebra  in $2+1$ dimensions is obtained by taking the limit $\epsilon \to 0$ of the $\mathfrak{su}(1,2)\oplus \mathfrak{u}(1)$ algebra to acquire \eqref{eq:sa}.

Next, we discuss the relation between the $\mathfrak{su}(1,2)\oplus \mathfrak{u}(1)$ algebra and the three-dimensional extended Schr\"odinger algebra \cite{Hartong:2016yrf, Kasikci:2020qsj}.\,\footnote{See related discussions for various extensions of (supersymmetric) Bargmann algebra~\cite{Papageorgiou:2009zc, Bergshoeff:2016lwr} and Newton-Hooke algebra~\cite{Papageorgiou:2010ud,Hartong:2017bwq}.} 
The latter algebra arises from the construction of
an off-shell Newton-Cartan gravity using three dimensional Chern-Simons theory.
The extended Schr\"odinger algebra reads \cite{Hartong:2016yrf}
\begin{subequations} \label{eq:GalileanA}
\begin{align}
    [H\,,\,G_i] &= P_i\,, 
        &%
    [P_i\,,\,C] &= G_i\,, 
        &%
    [H\,,\,C] &= D, 
        &%
    [P_i\,,\,G_j] &= \delta_{ij} \, N_0 - \epsilon_{ij} \, Y, \\[4pt]
    [D\,,\,H] &= - 2 \, H,
        &%
    [D\,,\,C] &= 2 \, C, 
        &%
    [D\,,\,P_i] &= -P_i\,,
        &%
    [D\,, \, G_i] &= G_i\,, \\[4pt]
    [J\,,\,P_i] &=\epsilon_{ij} \, P_j\,, 
        &%
    [J\,,\,G_i] &=\epsilon_{ij} \, G_j\,, \\[8pt]
    [P_i\,, \, P_j] &= \epsilon_{ij} \, Z ,
        &%
    [G_i\,, \, G_j]&= \epsilon_{ij} \, S, 
        &%
    [H \,, \, S] &= -2 \, Y,
        &%
    [H \,,Y]&=-Z, \\[4pt]
    [C \,, \, Y]&=S,
        &%
    [C \,, \, Z]&=2 \, Y,
        &%
    [D \,, \, S]&= 2 \, S,
        &%
    [D\,, \, Z]&= - 2 \, Z, 
\end{align}
\end{subequations}
where three extra generators $S,Y,Z$ are introduced compared to the (standard) Schr\"odinger algebra \eqref{eq:sa}.
In the limit where $S,Y,Z\to 0$, it reduces to the Schr\"odinger algebra. 
On the other hand, by grading the sl$(2,\mathbb{R})\oplus \mathfrak{u}(1)$ subalgebra in $\mathfrak{su}(1,2) \oplus \mathfrak{u}(1)$ via  
\beq
    T_I^{(n)} = T_I \otimes \epsilon^{n} \, ,
        \qquad%
    T_I=\{H,D,K,N\}\,,
\eeq 
we identify the generators in the extended Schr\"odinger algebra~\eqref{eq:GalileanA} with the generators in the $\mathfrak{su}(1,2) \oplus \mathfrak{u}(1)$ algebra~\eqref{eq:msuot} and \eqref{eq:msuot2} via
\begin{align}\label{eq:mapsbetweenalgebra}
\begin{split}
    Z = H \otimes \epsilon\,, 
        \quad%
    Y = - D \otimes \epsilon\,, 
        \quad%
    S = 2 \, C \otimes \epsilon\,, 
        \quad%
    N_0 = - 3 \, \Bigl( N + \,J \Bigr)  \otimes \epsilon \, .
\end{split}
\end{align}
This grading is interpreted as a large speed of light expansion in~\cite{Hansen:2019vqf,Hansen:2020pqs}.

%% file: Sections/App_details_map.tex
\section{\texorpdfstring{$\Omega$}{Omega}-Deformed Null Reduction}
\label{app:ssec:coord_ope}

In this appendix, we present how to obtain a TNC geometry with $\mathrm{SU}(1,2) \times \mathrm{U}(1)$ conformal symmetry in the operator picture by performing a null-reduction of Minkowski spacetime with an additional $\Omega$-deformation.
Our derivation is inspired from reference \cite{Lambert:2021nol}, but with a difference in the assignments of the length dimensions of the geometric data.
In \cite{Lambert:2021nol}, the matrix $\Omega_{ij}$ (that we are going to introduce below) has the dimension of an inverse length, thus introducing a length scale in the background.
This is at odds with having a background for the operator picture that does not include a length scale, \eg like Minkowski space.
To solve this issue, we will get rid of this length scale by making the time coordinate $T$ of dimension length squared, which fits with a background having $z =2$ Lifshitz scaling.

A useful starting point is the $\Omega$--deformed background
\begin{equation}
\dd s^2 =   2 \, \dd u \left(\dd T +  \Omega_{ij} \, x^j \, \dd x^i \right) + \dd x_i \, \dd x^i \, ,
\label{eq:ope0_app}
\end{equation}
where $i=1=1, \dots, 2(n-1)$.
This is a geometry with SU(1,$\, n$) conformal isometry obtained from a particular null-reduction of Minkowski space \cite{Lambert:2021nol}.
We briefly review how to find this metric.\footnote{While the manipulations below are similar to reference \cite{Lambert:2021nol}, we stress again that the coordinates that we introduce in Eq.~\eqref{eq:coord_transf_Lambert_ourconv} have a different length dimension, as we summarize in Eq.~\eqref{eq:scaling_Minkowski_Lambert}. }
Starting from Minkowski space
\begin{equation}
\dd s^2 = - 2 \, \dd \hat{x}^+ \, \dd \hat{x}^- + \dd \hat{x}^i \, \dd \hat{x}^i \, ,
\label{eq:lightcone_Minkowski}
\end{equation}
we introduce an arbitrary length scale $L$ and we define the coordinate transformation
\begin{equation}
\begin{aligned}
& \hat{x}^+ = - L \, \tan u \, , & \\
& \hat{x}^- = \frac{1}{L} \left( T - \frac{1}{2} \,  x^i \, x^i \,   \tan u \right) \, , & \\
& \hat{x}^i = x^i +  \Omega_{ij} \, x^j \, \tan u \, , &
\end{aligned}
\label{eq:coord_transf_Lambert_ourconv}
\end{equation}
with 
\begin{equation}
    \Omega_{ij} = - \Omega_{ji}\,, 
        \qquad%
    \Omega_{ij} \, \Omega_{ik} = \delta_{jk} \, .
\end{equation}
In these conventions, the length dimensions of the various objects are\footnote{For comparison, the length dimensions used in Eq.~(2.2) of reference \cite{Lambert:2021nol} read
\beq
 [\hat{x}^+] = [\hat{x}^-] = [\hat{x}^i] = 1 \, , \qquad
[R] = 1 \, , \qquad  
[\Omega] = -1 \, , \qquad
[x^+] = 1 \, , \qquad
[x^-]= 1 \, .   
\label{eq:scaling_Minkowski_Lambert_conv}
\eeq
}
\beq
 [\hat{x}^+] = [\hat{x}^-] = [\hat{x}^i] = 1 \, , \qquad
[L] = 1 \, , \qquad  
[\Omega] = 0 \, , \qquad
[u] = 0 \, , \qquad
[T]=2 \, .   
\label{eq:scaling_Minkowski_Lambert}
\eeq
The metric in the new coordinate system reads
\begin{equation}
\label{metric_before}
\dd s^2 = \frac{1}{\cos^2 u}  \Big[  2 \, \dd u \left(\dd T +\Omega_{ij} \, x^j \, \dd x^i \right) + \dd x_i \, \dd x^i\Big] \, .
\end{equation}
After performing a Weyl rescaling to get rid of the overall prefactor, one gets Eq.~\eqref{eq:ope0_app}.
We notice that any dependence on the length scale $L$ automatically disappeared from the metric due to its scale invariance.

Choosing $\Omega_{yx}=- \Omega_{xy}=1$, we obtain the metric in Cartesian coordinates
\beq
\dd s^2 = 2 \le \dd T - y \, \dd x + x \, \dd y \ri \dd u + \dd x^2 + \dd y^2 \, .
\label{eq:ope1_app}
\eeq
This background fits the null reduction form \eqref{eq:nullredmetric} once we identify
\beq
\tau = \dd T - y \, \dd x + x \dd \, y \, , \qquad
m=0 \, , \qquad
h_{\mu\nu} \, \dd x^{\mu} \, \dd x^{\nu} = \dd x^2 + 
\dd y^2 \, .
\label{eq:app:TNC_data}
\eeq
The TNC geometry with data \eqref{eq:app:TNC_data} describes a spacetime with $\mathrm{SU}(1,2) \times \mathrm{U}(1)$ conformal isometry.